\documentclass[traditabstract]{aa} 
\usepackage{graphicx}
\usepackage{txfonts}
\newcommand{\solm}{M$_{\odot}$\ }
\newcommand{\solar}{L$_{\odot}$\ }

\newcommand{\rf}{\par\noindent\hangindent 15pt {}}

\begin{document}

\authorrunning{Sabha, Witzel, Eckart et al.}  
\titlerunning{The extreme luminosity states of SgrA*}
\title{The extreme luminosity states of Sagittarius~A*}
\subtitle{}
   \author{N. Sabha\inst{1}
          \and
          G. Witzel\inst{1}
          \and 
          A. Eckart\inst{1,2}
          \and 
          R.M. Buchholz\inst{1}
	  \and
          M. Bremer\inst{1}
	  \and
          R. Gie\ss \"ubel\inst{2,1}
	  \and
          M. Garc\'{\i}a-Mar\'{\i}n\inst{1}
          \and 
          D. Kunneriath\inst{1,2}
	  \and
          K. Muzic\inst{1}
	  \and
	  R. Sch\"odel\inst{3} 
	  \and
          C. Straubmeier\inst{1}
	  \and
          M. Zamaninasab\inst{2,1} 
	  \and
          A. Zernickel\inst{1} 
          }
\offprints{N. Sabha (sabha@ph1.uni-koeln.de)}

   \institute{ I.Physikalisches Institut, Universit\"at zu K\"oln,
              Z\"ulpicher Str.77, 50937 K\"oln, Germany\\
              \email{sabha, witzel, eckart@ph1.uni-koeln.de}
         \and
             Max-Planck-Institut f\"ur Radioastronomie, 
             Auf dem H\"ugel 69, 53121 Bonn, Germany
         \and
             Instituto de Astrof\'isica de Andaluc\'ia (CSIC), 
	     Camino Bajo de Hu\'etor 50, 18008 Granada, Spain\\
              \email{rainer@iaa.es}
             }

\date{Received  / Accepted }

\abstract{
We discuss mm-wavelength radio,  2.2-11.8$\mu$m NIR and 
2-10~keV X-ray light curves of the super massive black hole (SMBH) 
counterpart of Sagittarius A* (SgrA*) near its lowest and 
highest observed luminosity states.
We investigate the structure and brightness of the central 
S-star cluster harboring the SMBH to obtain reliable flux density estimates of SgrA* 
during its low luminosity phases.
We then discuss the physical processes responsible for 
the brightest flare as well as the faintest flare or quiescent 
emission in the NIR and X-ray domain.
To investigate the low state of SgrA*
we use three independent methods to remove or strongly  suppress the
flux density contributions of stars in the central 2'' diameter region 
around SgrA*. 
The three methods are: a) low-pass filtering the image;
b) iterative identification and removal of individual stars;
c) automatic point spread function (PSF) subtraction.
For the lowest observed flux density state all 3 image reduction methods result in 
the detection of faint
extended emission with a diameter of 0.5'' - 1.0'' and
centered on the position of SgrA*. 
We analyzed two datasets that cover the lowest luminosity states of SgrA* we observed to date. 
In one case we detect a faint K-band (2.2$\mu$m)
source of $\sim$4~mJy brightness (de-reddened with $A_K$=2.8) which we identify as SgrA* in its low state. 
In the other case no source brighter or equal to a de-reddened 
K-band flux density of 
$\sim$2~mJy was detected at that position.
As physical emission mechanisms for SgrA* we discuss bremsstrahlung, 
thermal emission of a hypothetical optically thick disk,
synchrotron and synchrotron self-Compton (SSC) emission,
and in the case of a bright flare the associated radio response due to 
adiabatic expansion of the synchrotron radiation emitting source component.
The luminosity during the low state can be interpreted as 
synchrotron emission from a continuous or even spotted accretion disk.
For the high luminosity state SSC emission from THz peaked source components
can fully account for the flux density variations observed in the
NIR and X-ray domain.
We conclude that at near-infrared wavelengths the SSC mechanism
is responsible for all emission from the lowest to 
the brightest flare from SgrA*.
For the bright flare event of 4 April 2007 that was covered from the radio to the X-ray 
domain, the SSC model combined with adiabatic expansion can explain the related peak
luminosities and different widths of the flare profiles obtained in 
the NIR and X-ray regime
as well as the non detection in the radio domain.
}


\titlerunning{The Extreme Luminosity States of SgrA*}
\authorrunning{Sabha, Witzel, Eckart et al.}  
\maketitle
%

\section{Introduction}
\label{section:Introduction}


\begin{table*}
\begin{center}
\caption{The brightest X-ray flares from SgrA*. For references see text.
\label{strongflares}}
{\begin{small}
\begin{tabular}{ccccclccc} \hline
label & date & Observatory & $\alpha=\Gamma-1$ & X-ray flux      &NIR L-band  & 11.8$\mu$m   \\
 &    &             &     &  density        &flux density& flux density \\
 &    &             &     & ($\mu$Jy)       & (mJy)      &  (mJy)       \\
\hline
$\alpha$ & Oct. 26/27, 2000 & Chandra & 0.0$\pm$0.8 & 0.65$\pm$0.07 &     -     & -     \\
$\beta$  & Oct. 3, 2002     & XMM     & 1.2$\pm$0.3 & 2.7$\pm$0.4   &     -     & -     \\
$\gamma$ & Apr. 4, 2007     & XMM     & 1.3$\pm$0.3 & 1.7$\pm$0.3   & 20$\pm$2  & $<$3$\sigma$=57 \\
\hline
\end{tabular}
\end{small}}
\end{center}
\end{table*}


\begin{table*}
\centering
\caption{Details on observing runs during which SgrA* was in a very low NIR luminosity state.
\label{weakflares}}
{\begin{small}
\begin{tabular}{cccll}
\hline
Telescope & Instrument & $\lambda$ & UT and JD & UT and JD \\
Observing ID & & & Start Time & Stop Time \\
\hline
VLT UT~4       & NACO      & 2.2~$\mu$m  & 2004 23 Sept. 23:20:45 & 2004 24 Sept. 01:45:11 \\
               &           &             & JD 2453272.47969 & JD 2453272.57304\\
VLT UT~4       & NACO      & 2.2~$\mu$m  & 2004 30 Aug. 23:49:36 &  2004 31 Aug. 01:28:19 \\
               &           &             & JD 2453248.48375 & JD 2454466.50000 \\
\hline
\end{tabular}
\end{small}}
\end{table*}


At the center of the Milky Way 
stellar motions and variable emission allow us to firmly 
associate Sagittarius A* 
(SgrA*) with a 4$\times$10$^6$ \solm
super-massive black hole 
(Eckart \& Genzel 1996, Genzel et al. 1997, 2000, Ghez et al. 1998, 2000, 2003, 2005a, 2008, 
Eckart et al. 2002, Sch\"odel et al. 2002, 2003, 2009, Eisenhauer 2003, 2005, Gillessen et al. 2009).

Recent radio, near-infrared and X-ray observations have detected variable 
and polarized emission and give detailed insight into the 
physical emission mechanisms at work in SgrA*,which may include any or all of synchrotron, SSC, 
and bremsstrahlung emission
(e.g. Baganoff et al. 2001, 2002, 2003, 
Eckart et al. 2003, 2004, 2006ab, 2008ab, 2009,
Porquet et al. 2003, 2008, Goldwurm et al. 2003, Genzel et al. 2003, 
Ghez et al. 2004ab, Eisenhauer et al. 2005, 
Belanger et al. 2006, Hornstein et al. 2007,
Yusef-Zadeh et al. 2006ab, 2007, 2008, 2009, Marrone et al. 2009,  
Dodds-Eden et al. 2009).
Multi wavelength detections of the radio point source 
at sub-millimeter, X-ray, and infrared wavelengths have 
also been made, showing that the luminosity associated 
with SgrA* is on the order of
10$^{-9...-10}$ times
below the Eddington luminosity L$_{Edd}$ and many orders of magnitudes below
that of SMBHs in active galactic nuclei (AGN) with 
comparable masses (of about 4$\times$10$^6$\solm). 
The surprisingly low luminosity has motivated many theoretical 
and observational efforts to explain the processes that are at work 
in Sgr A*.
For a recent summary of accretion models and variable accretion 
of stellar winds onto Sgr~A* see Yuan (2006) and Cuadra \& Nayakshin (2006, 2009).

Sgr~A* is - in terms of Eddington luminosity - the faintest super-massive black hole
known.
However, due to its proximity it is bright enough to be studied in great detail. 
With the possible exception of the closest galaxies,
no extragalactic super-massive black hole with a similar feeble Eddington rate would
be observable. 
Motivated by the need to explain the variable luminosity of SgrA*
we have analyzed data from it highest and lowest states.
In order to separate the weak NIR emission of Sgr~A* from the surrounding 
stars, the use of large telescopes and adaptive optics (AO) 
is required. 
Here we present the first attempt to do this with near-infrared data obtained with the VLT.
Likewise, high resolution is needed in the X-ray regime to separate 
Sgr~A* from the surrounding diffuse X-ray background.

The temporal correlation between
the rapid variability of the near-infrared (NIR) and X-ray emission
suggests that the emission showing 10$^{33-35}$~erg/s flares
arises from a compact source within a few tens of 
Schwarzschild radii of the SMBH.
For SgrA* we assume 
$R_s$=2$R_g$=2GM/c$^2$$\sim$10$^{10}$~$m$
for a $\sim$4$\times$10$^6$\solm ~black hole. 
Here $R_s$ is one 
Schwarzschild radius and $R_g$ the gravitational radius of the SMBH.
One R$_s$ then corresponds
to an angular diameter of $\sim$8~$\mu$as at a distance to the Galactic
center of $d=8~kpc$ 
(Reid 1993, Eisenhauer et al. 2003, Ghez et al. 2005ab).
These observations can be explained in a model of 
an intrinsically faint accretion disk that is dominated by red-noise
and temporarily harbors 
a bright orbiting spot possibly in conjunction with a short jet 
(Eckart et al. 2006b, Meyer et al. 2006ab, 2007).
The phenomenon can best be studied through the brightest flares
that allow us to derive high signal to noise light curves and 
multi wavelength observations throughout the electromagnetic spectrum.

In its lowest luminosity states SgrA* is difficult to detect though 
due to the presence of a strong diffuse background
at X-ray wavelengths and the confusion with nearby stellar sources 
in the near-infrared. 
Here we discuss the relative importance of the bremsstrahlung and
SSC process for the low state of SgrA* and investigate if the
brightest flares of SgrA* can be explained via 
a SSC model involving up-scattered
sub-millimeter photons from compact source components.
In Sects.~\ref{section:Observations} and
\ref{section:SubtractingStars} we describe the observations, the
data reduction and the algorithms we used to extract SgrA* in its low 
state from the background emission of the dense cluster of high velocity 
stars at angular separations of less than about 1'' from the position of SgrA*. 
Below we will refer to this association of stars as the
'S-star cluster'.
In Sect.~\ref{section:ResultsandDiscussion}
we present and discuss results.
We start by summarizing the results of our deep Ks imaging
in subsections \ref{subsection:ImagingTheCusp} and
\ref{subsection:sgralow}.
In the following subsections 
we then discuss various emission mechanisms
for the low and high luminosity states of SgrA*.
We show that the SSC process can successfully discribe both the low and 
high luminosity states and that - in particular for the high states - no
extra mechanisms need to be involved (discussion in e.g.
Dodds-Eden et al. 2009, Eckart et al. 2009).
A brief summary and conclusions are presented in Sect.~\ref{section:Summary}.

Throughout the paper we quote de-reddened infrared flux densities 
using a Ks-band (2.2$\mu$m) extinction correction of $A_{Ks} = 2.8$
(Scoville et al. 2003,
Sch\"odel et al. 2007,
Sch\"odel et al. 2010 in prep.,
Buchholz, Schoedel, Eckart 2009).
In Sect.~\ref{subsection:ImagingTheCusp}
we use $\Gamma$ and $\gamma$ to describe the projected and three 
dimensional structure of the central stellar cluster respectively. 
In Table~\ref{strongflares} and  
starting with subsection~\ref{subsection:SSC}
we use $\Gamma$ and $\gamma$ 
as the bulk boosting factor and the relativistic
electron Lorentz factor.

\noindent
\begin{figure*}
\centering
\includegraphics[width=18cm,angle=-00]{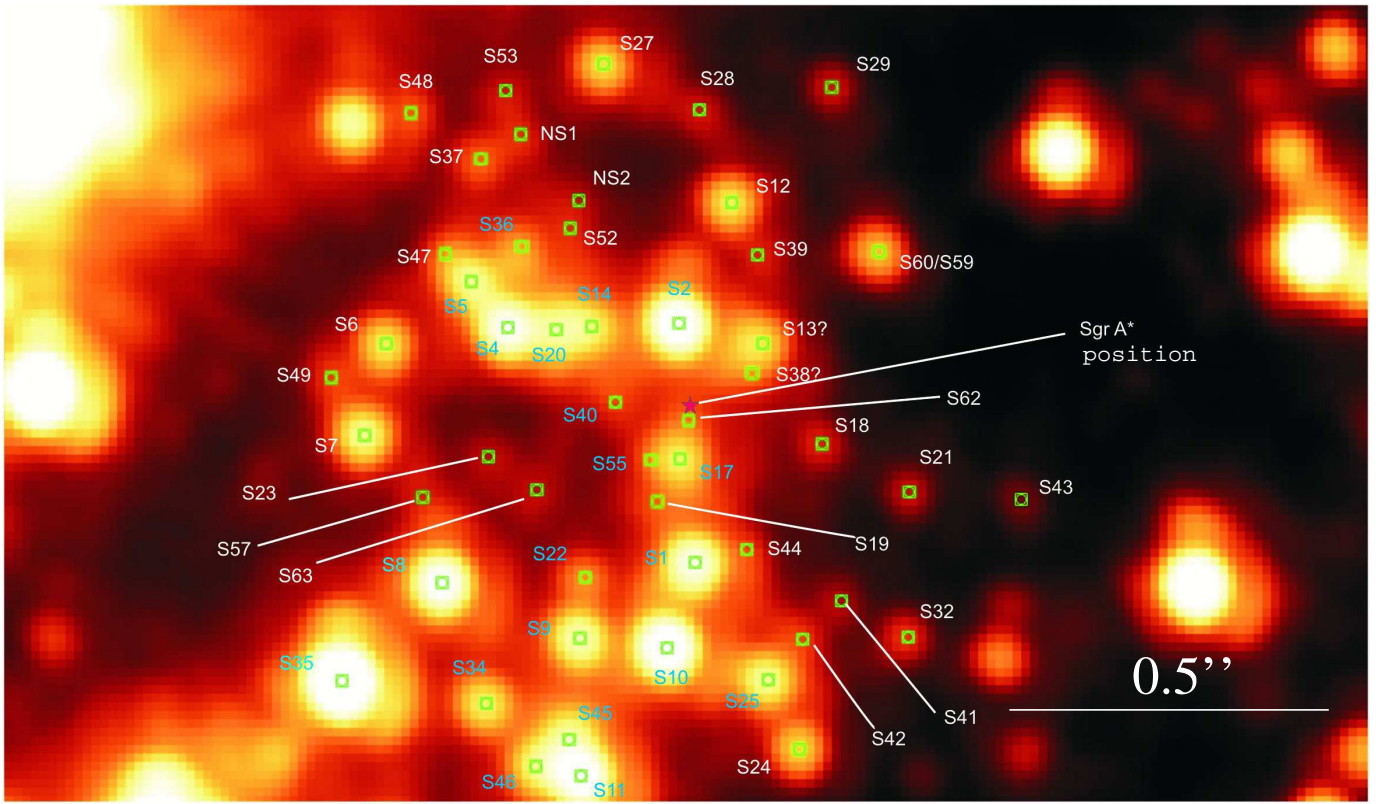}
\caption{\small
Identification of individual sources that were used for the iterative
subtraction of the Ks-band datasets (Table~\ref{weakflares}).
Here we show the image derived from the 23 September 2004 data.
The nomenclature was taken from the deconvolved H-band image given in Fig.1 by 
Gillessen et al. (2006). 
Sources that are not contained in there have labels starting with 'N'.
Relative positions and flux densities of the labeled sources are given in 
Table~\ref{list}.
}
\label{fig-0}    
\end{figure*}



\begin{table*}[!htp]
\begin{center}
\caption{Sources used for iterative source subtraction for the
 23 September 2004 data.
\label{list}}
{\begin{small}
\begin{tabular}{llrrrrrrr} \hline
name & other  & $\Delta$$\alpha$ & $\Delta$$\delta$ &   $R$  & flux & Ks & K'-0.2 & $\Delta$mag\\
     & name   &  arcsec          &  acrsec          & arcsec & mJy  &    &    &   \\
\hline
S62   &      &   0.01&  -0.02&  0.03&  1.52&  17.0&      &      \\
S17   &S0-17 &   0.02&  -0.09&  0.09&  8.06&  15.2&  15.5&  -0.3\\
S38?  &      &  -0.08&   0.05&  0.10&  1.27&  17.2&      &      \\
S55   &      &   0.07&  -0.09&  0.11&  0.71&  17.9&      &      \\
S40   &      &   0.12&   0.00&  0.12&  1.88&  16.8&      &      \\
S2    &S0-2  &   0.02&   0.13&  0.13& 22.00&  14.1&  13.9&   0.2\\
S13?  &S0-8  &  -0.11&   0.09&  0.14&  7.04&  15.4&    ? &      \\
S19   &      &   0.06&  -0.15&  0.16&  1.79&  16.9&      &      \\
S14   &      &   0.16&   0.12&  0.20&  7.41&  15.3&      &      \\
S18   &S0-38 &  -0.20&  -0.06&  0.21&  2.19&  16.7&    ? &      \\
S44   &      &  -0.08&  -0.23&  0.24&  1.13&  17.4&      &      \\
S20   &      &   0.21&   0.12&  0.24&  8.80&  15.1&      &      \\
S39   &      &  -0.10&   0.23&  0.25&  1.37&  17.2&      &      \\
S1    &S0-1  &   0.00&  -0.25&  0.25& 14.00&  14.6&  14.5&   0.1\\
S63   &      &   0.24&  -0.13&  0.28&  1.20&  17.3&      &      \\
S4    &S0-3  &   0.29&   0.12&  0.31& 14.02&  14.6&  14.3&   0.3\\
S22   &      &   0.17&  -0.27&  0.32&  2.39&  16.6&      &      \\
S12   &      &  -0.06&   0.31&  0.32&  8.04&  15.2&      &      \\
S23   &      &   0.32&  -0.08&  0.33&  1.01&  17.5&      &      \\
S52   &      &   0.19&   0.27&  0.33&  0.98&  17.5&      &      \\
NS2   &      &   0.18&   0.32&  0.36&  1.06&  17.4&      &      \\
S21   &      &  -0.33&  -0.14&  0.36&  2.03&  16.7&      &      \\
S36   &      &   0.27&   0.25&  0.36&  3.66&  16.1&      &      \\
S60/59&S0-12 &  -0.29&   0.24&  0.37&  6.70&  15.4&  14.2&   1.2\\
S41   &      &  -0.23&  -0.31&  0.38&  1.10&  17.4&      &      \\
S10   &S0-6  &   0.04&  -0.38&  0.38& 21.99&  14.1&  14.0&   0.1\\
S5    &      &   0.34&   0.19&  0.39&  9.47&  15.1&      &      \\
S42   &      &  -0.17&  -0.37&  0.40&  1.62&  17.0&      &      \\
S9    &S0-5  &   0.18&  -0.36&  0.40& 12.66&  14.7&  14.9&  -0.2\\
S57   &      &   0.42&  -0.14&  0.44&  0.89&  17.6&      &      \\
S25   &S0-18 &  -0.11&  -0.43&  0.44& 10.71&  14.9&   ?  &      \\
S47   &      &   0.38&   0.23&  0.45&  2.73&  16.4&      &      \\
S28   &      &  -0.01&   0.46&  0.46&  1.56&  17.0&      &      \\
S6    &      &   0.47&   0.09&  0.48&  8.02&  15.2&      &      \\
S8    &S0-4  &   0.39&  -0.28&  0.48& 17.72&  14.4&  14.3&   0.1\\
S32   &      &   0.33&  -0.36&  0.49&  2.85&  16.4&      &      \\
NS1   &      &   0.27&   0.42&  0.50&  1.81&  16.9&      &      \\
S37   &      &   0.33&   0.38&  0.50&  2.74&  16.4&      &      \\
S7    &S0-11 &   0.51&  -0.05&  0.51& 10.38&  15.0&  15.1&  -0.1\\
S43   &      &  -0.51&  -0.15&  0.53&  1.30&  17.2&      &      \\
S29   &      &  -0.21&   0.49&  0.54&  2.33&  16.6&      &      \\
S27   &S0-27 &   0.14&   0.53&  0.55&  7.34&  15.3&  15.5&  -0.2\\
S49   &      &   0.56&   0.04&  0.56&  1.60&  17.0&      &      \\
S45   &      &   0.19&  -0.52&  0.56&  7.59&  15.3&      &      \\
S24   &S0-28 &  -0.16&  -0.54&  0.56&  6.26&  15.5&  15.6&  -0.1\\
S53   &      &   0.29&   0.49&  0.57&  1.30&  17.2&      &      \\
S34   &      &   0.32&  -0.47&  0.57&  8.55&  15.2&      &      \\
S11   &S0-9  &   0.17&  -0.58&  0.60& 23.79&  14.1&  14.2&  -0.1\\
S46   &      &   0.24&  -0.56&  0.61&  6.92&  15.4&      &      \\
S48   &      &   0.44&   0.45&  0.63&  3.21&  16.2&      &      \\
S35   &S0-13 &   0.54&  -0.43&  0.69& 45.90&  13.3&  13.3&   0.0\\
\hline
\end{tabular}
\end{small}}
\end{center}
\end{table*}


\section{Observations and data reduction}
\label{section:Observations}

Near-infrared (NIR) observations of the Galactic center (GC) were
carried out with the NIR camera CONICA and the adaptive optics (AO)
module NAOS (briefly ``NACO'') at the ESO VLT unit telescope~4 (YEPUN) 
on Paranal, Chile. 
The data were taken on the night of 30 August and 23 September, 2004. 
The dataset taken on 23 September 2004
turned out to be the best available set in which SgrA* 
is in a quiet state. 
During the 30 August observations some weak activity at 2.2$\mu$m could be
detected, making this dataset well-suited to determine 
the position of SgrA* with respect to neighboring stars.
The combination of both datasets is ideally suited to obtain a
flux density estimate or limit of SgrA* in its quiet state.

For all observations the infrared wavefront sensor of NAOS was used 
to lock the AO loop on the NIR bright (Ks-band magnitude $\sim$6.5) 
supergiant IRS~7, located about $5.6''$ north of Sgr~A*.  
The pixel scale was 13.27~mas.

The atmospheric conditions (and consequently the AO correction) were
stable during the observations. The seeing at the telescope measured in the optical was $\sim$0.6''.
We used an integration time of DIT=2~seconds and a number of integrated images of NDIT=15.
Details on start and stop times are listed in Table~\ref{weakflares}.

All observations in the Ks-band (2.2~$\mu$m)
were dithered to cover a larger area of the GC by mosaic imaging. 
The sky background for the Ks-band observations was extracted from the median of
stacks of dithered exposures
of a dark cloud -- a region practically empty of stars --
about 400'' north and 713'' west of the target. 
All exposures were sky-subtracted, flat-fielded, and corrected for
dead or bad pixels. 
The images were shifted and stacked 
in a cube with a mean average to get 
a mosaic image of the GC. All of these steps 
were performed with the DPUSER software for
astronomical image analysis (T. Ott, MPE; see also Eckart \& Duhoux 1990).
Subsequently, PSFs were extracted from these
images with \emph{StarFinder}
(Diolaiti et al. 2000) both for deconvolution 
with the Lucy-Richardson (LR Lucy 1974) algorithm and for PSF subtraction.

The flux densities of the sources were measured by aperture photometry with
circular apertures of 40~mas radius and corrected
for extinction, using $A_{K} = 2.8$
(Scoville et al. 2003,
Sch\"odel et al. 2007,
Sch\"odel et al. 2010 in prep.,
Buchholz, Schoedel, Eckart 2009).
Possible uncertainties in the extinction by a few tenths of a magnitude 
alter the infrared flux densities of SgrA* by a few 10\%. These variations 
are easily compensated for by insignificant variations in the spectral indices
or the boosting which do not influence the general results obtained in this paper.

The flux density calibration was carried out with the
known Ks-band flux densities of IRS16C, IRS16NE, and IRS21 by 
Blum, Sellgren \& Depoy (1996).
Subsequently the relative photometry for Sgr~A*
was done using 10 sources within 1\farcs6 of Sgr~A* as secondary calibrators
(S67, S92, S35, S8, S76, S1, S2, S87, S65, S30; Gillessen et al. 2006).  
This results in a Ks-band flux density of the high velocity star
S2 of 22$\pm$1~mJy, which 
compares well with the magnitude and flux for S2 quoted by 
Ghez et al. (2005b) and Genzel et al. (2003).
The measurement uncertainties for Sgr~A* were obtained on the nearby reference star S2.
The background flux in the immediate vicinity of Sgr~A* was 
obtained by averaging the measurements at six random locations 
in a field located about 0\farcs6 west of Sgr~A* that is
free of obvious stellar sources.
A Ks-band image of the central 4.3''$\times$2.5'' is shown in Fig.~\ref{fig-0}.
This field is at the center of the stellar cluster of which we show
a 15''$\times$15'' overview in Fig.~\ref{fig-1}a).
The source positions and relative astrometry was established
via \emph{StarFinder}. 
The positions in 
Table~\ref{list} are given relative to the position of 
SgrA* as determined from the 23 September 2004 data.
We list the name following the nomenclature used by Gillessen et al. (2009) 
and in the second column the name used by Do et al. (2009b). 
Sources labeled with $N$ had not been identified before.
Identifications with question marks are tentative and suffer most likely 
from the proper motion of faint sources between the present date and the date 
used for identification by Gillessen et al. (2009).
Then we list in Table~\ref{list} the sky projected relative offsets
 $\Delta$$\alpha$, $\Delta$$\delta$ and radial distance $R$ from SgrA* in arcseconds.
The uncertainties in the offsets are 
better than a third of a pixel i.e. 0.004''
We then list in Table~\ref{list}
the observed Ks magnitude derived in this paper
and the observed K' magnitude listed by Do et al. (2009b) 
corrected by an offset value of -0.2 magnitudes.
The scatter between both values after correction for this offset is 0.2 magnitudes, so that the
offset corresponds to 1$\sigma$ in magnitudes (20\% in flux)
and is well within the expected calibration uncertainties for faint 
stars in the crowded field.
The difference between both mangitude values is listed in the last column.
For sources S0-8, S0-18 and S0-38 no magnitudes are given by Do et al. (2009b).
The large magnitude difference in the case of S0-12 is most
likely due to the blend of S59 and S60. 
The two NIR filters $K_s$ and K' (Do et al. 2009b) are very similar with 
Ks(NACO): $\lambda_{center}$=2.18$\mu$m and width $\Delta$$\lambda$=0.35$\mu$m
and
K`(NIRC): $\lambda_{center}$=2.124$\mu$m and width $\Delta$$\lambda$=0.351$\mu$m.
No transformation between magnitudes derived with these filters was applied.
The 0.2~mag scatter relates to
stars of different brightnesses at different epochs between which they are
moving at different locations in the cluster with respect to different
local backgrounds.
The relative flux density uncertainty for the star S2 - which we use as an estimate for SgrA* -
is determined within a single epoch at fixed position.

\section{Subtracting stars}
\label{section:SubtractingStars}

To investigate the low luminosity state of SgrA*
we use three independent methods to remove or strongly suppress the
flux density contributions of stars in the central 2'' diameter region 
around SgrA*. 
All three methods give comparable results and allow us to clearly 
determine the stellar light background at the center of the Milky Way
against which SgrA* has to be detected.

We assume that the PSF as determined from stars in the central few arcseconds of the
image is uniform across the central S-star cluster.
Investigations of larger images
(e.g. Buchholz et al. 2009)  show that on scales of a few arcesconds  
this is a reasonable assumption and that 
PSF variations have to be taken into account
only for fields $\ge$10''.
Whether the residual extended emission we find close to SgrA* is due to 
rsiduals from the subtracting algorithms or associated with additional
stellar emission from the S-star cluster is discussed in 
Sect.~\ref{subsection:ImagingTheCusp}.

\subsection{Linear extraction of extended flux density}
\label{subsection:Linear}

With the point spread function $P$ and the object distribution $O$
we can write the formation of an image $I$ as

\begin{equation}
I = O * P = (\Delta + E) * P = \Delta * P + E * P~.
\end{equation}

Here the symbol $*$ denotes the convolution operator.
The variable $\Delta$ denotes the distribution of stellar sources 
and $E$ the distribution of extended emission.
Both types of sources are assumed to be ideally 
unresolved or resolved with respect to the 
point spread function.
To produce a high-pass filtered image we can calculate a 
smooth-subtracted image $\Sigma$ as

\begin{equation}
\Sigma = I*G - I = (\Delta + E) * P * G - (\Delta + E) * P~.
\end{equation}

Here we use a narrow Gaussian $G$ normalized to an integral power of unity. 
This can be rewritten as

\begin{equation}
\Sigma =  \Delta * (P * G - P) +  (E * P * G - E * P)~.
\end{equation}

Since the width of the extended emission is expected to be 
much larger than the full width at half maximum of the PSF,
and in particular since the Gaussian $G$ is narrow with respect to $P$,
we can infer that $E * P * G \sim  E * P$.
Then the differential point spread function 
$P * G - P $
is the PSF of the high-pass filtered differential image

\begin{equation}
\Sigma \sim  \Delta * (P * G - P)~. 
\end{equation}

Since $\Sigma$ is fully dominated by the contribution of the
distribution of unresolved stellar sources, this distribution
$\Delta$ can be retrieved via a linear deconvolution of the 
differential image with the differential PSF

\begin{equation}
\Delta  \sim  \Sigma~\%~(P * G - P)~. 
\end{equation}

Here the symbol $\%$ denotes the deconvolution operation.
Consequently the distribution of extended flux 
(i.e. a low-passed filtered version of the image)
$E * P  $ can now be isolated via

\begin{equation}
E * P = I - \Delta * P
\end{equation}
or
\begin{equation}
I - \Sigma * P~\%~(P * G - P)  =  
\Delta * P + E * P - \Delta * P  = E * P~.  
\end{equation}

\begin{figure}
\centering
\includegraphics[width=9cm,angle=-00]{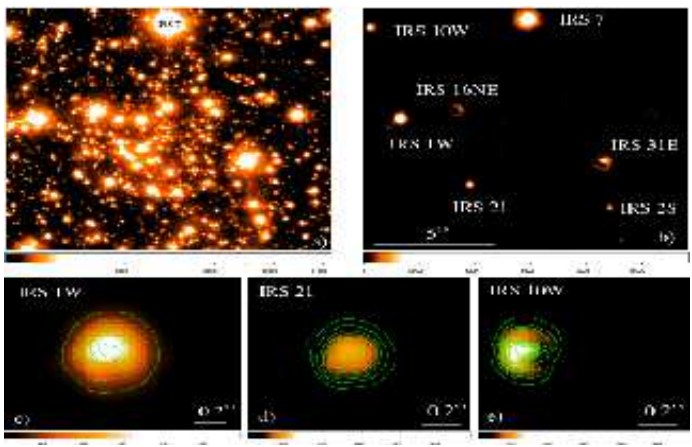}
\caption{\small
The Ks-band image of the Galactic center region
{\bf (a)} 
taken on 23 September 2004. The central
 15''$\times$15'' image was taken with the adaptive optics system 
NACO at the ESO VLT and the results of the low-pass filter algorithm.
{\bf (b,c,d,e)}. The contour lines are separated by 
8\% of the peak brightness of IRS10W.
}
\label{fig-1}    
\end{figure}

The advantage of this new
method is that it is independent
of input models, i.e. 
it does not depend on the identification of individual stars
or on assumptions on the existence or properties of a 
diffuse background flux density distribution.
Also the method only uses linear operations, i.e. image 
subtraction and linear convolution and deconvolution 
(Fourier multiplications and divisions).
Therefore it is not dependent on the 
flux density of sources and numbers of iterations, as  
is the case for many non-linear algorithms 
(e.g. the Lucy-Richardson or maximum entropy algorithm). 
In order to demonstrate the validity of the approach we show
the results we obtained 
on the central 15''$\times$ 15'' of the 23 September data
in Fig.~\ref{fig-1}.
For all the prominent extended Ks-band sources 
(Fig.~\ref{fig-1}) the extended Ks-band emission can be 
retrieved and its structure compares very well
to the previously published results (e.g. Ott, Eckart \& Genzel 1999,
Tanner et al. 2002, 2005).
(A comparison of different deconvolution algorithms used in the 
crowded Galactic center field is given in 
Ott, Eckart \& Genzel 1999.)
At a dynamic range of about 100:1 we show 
in Fig.~\ref{fig-1}b) a low-pass filtered version of the Ks-band image
that was obtained by following the algorithm outlined above.
The prominent sources have been labeled with their names.
In Figs.~\ref{fig-1}c), d) and e) we show the extended flux density 
distribution of the following sources: IRS~1W,  IRS~21,  IRS~10W.
These bow-shock structures are shown at the angular resolution of about 
0.1'' as delivered by the AO system. They compare very well with the 
source structures published by Tanner et al. 2005.
Here the algorithm detects very bright sources like IRS~7 also as an 
extended object since the star 
is overexposed and not well described by the used PSF.
In Fig.~\ref{fig-2} we show results of the algorithm 
on the central 2''$\times$ 2''. 
In addition to residuals of the filter process around the position of 
bright stars, we see an increased flux density level within the dashed 
0.5'' diameter circle centered on the position of SgrA*. 
The disadvantage of this algorithm is that any point-like 
emission from SgrA* is also strongly suppressed.
The residuals (ringing) still evident around the bright stars 
are a consequence of the choice of the apodization filter 
that needs to be defined in the case of linear deconvolution.
This filter is used to suppress the usual high spatial 
frequency noise that arises due to small number division problems.
We used a cosine-bell shaped filter that preserves the high
angular resolution but results in some residual ringing.
However, the result on the background emission is not
strongly influenced by this, since the bright stars are
sufficiently far away from the central position.
The comparison to the results of the other two
algorithms used to remove
the contributions of stars near SgrA* shows that the result
is also little affected by the much fainter stars
close to SgrA*. Their contribution has been efficiently
suppressed by the algorithm.

\subsection{Iterative PSF subtraction}
\label{subsection:Iterative}

The iterative PSF subtraction was carried out within 
the central 2'' to 3'' around the position of SgrA*.
This was performed by resizing, shifting and scaling 
the PSF to the position and the value of each 
star and then subtract it.
The sources used for this process are listed in 
Table~\ref{list} and  are shown in Fig.~\ref{fig-0}. 
A total number of 51 
stars were subtracted so that the resulting 
background was as smooth as possible. 
In Fig.~\ref{fig-3} we show the resulting subtracted image
of the area of interest around SgrA*. 
All stars brighter than about 17.5$^m$ (see KLF in Fig.~\ref{fig-5})
have been subtracted with a scaled version of the PSF.
The thin lines in Figs.~\ref{fig-3}a) and b) are used to 
triangulate the position of SgrA*.
Fig.~\ref{fig-3}a) shows the image obtained from a section 
of the 30 August dataset with SgrA* flaring.  
Compared to S2 with $\sim$22~mJy SgrA* has a brightness of about 4~mJy. 
Fig.~\ref{fig-3}b) shows the image obtained from a section of the 
23 September dataset with SgrA* in a quiescent phase.  
Here SgrA* is clearly not detected and is fainter than 
about 2~mJy (de-reddened).

Figure~\ref{fig-3} also reveals the presence of faint residual
extended emission - centered on the position of SgrA*.
This residual emission is most likely from the 
central S-star cluster of high velocity stars.
We determined the position of SgrA* from the data with 
SgrA* in a flare state taken on 30 August, 2004. 
Taking an upper limit of the observed two dimensional projected 
stellar velocity dispersion of $\sim$1000~km/s and the fact that
for the Galactic center proper motion of 1~mas/yr corresponds to
about 39~km/s, the relative positional uncertainty for stars observed 
between both epochs is expected to be well below 2~mas.
Thin lines in Figs.~\ref{fig-3}a) and b) are used to 
triangulate the position of SgrA*.
Figure~\ref{fig-3} shows that for the dataset taken
on 23 September 2004 no NIR counterpart of SgrA* was detected.
For 30 August 2004 we find a NIR counterpart of SgrA* with a
dereddened flux density of about 4~mJy with a 3$\sigma$ uncertainty of about 2~mJy.
Estimates of the flux density limits and uncertainties are given in 
Sect.~\ref{subsection:sgralow}.

\subsection{Automatic PSF subtraction}
\label{subsection:automatic}

\begin{figure}
\centering
\includegraphics[width=9cm,angle=-00]{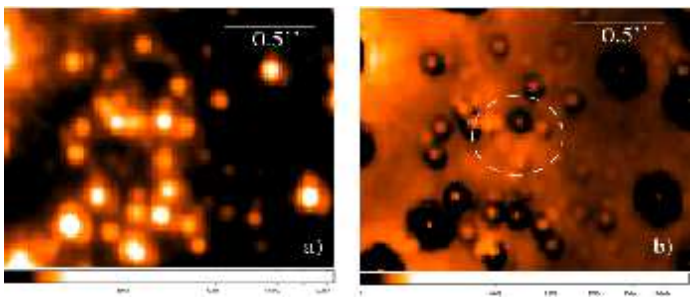}
\caption{\small
The inner 2''$\times$2'' of the central stellar cluster
(23 September 2004). 
The position of SgrA* is at the center of the images a) and b).
In panel a) we show the corresponding image section taken out 
of the Ks-band image in Fig.~\ref{fig-1}~a).
In panel b) we show the corresponding image section taken out 
of the low-pass filtered version in Fig.~\ref{fig-1}~b).
}
\label{fig-2}    
\end{figure}

\begin{figure}
\centering
\includegraphics[width=9cm,angle=-00]{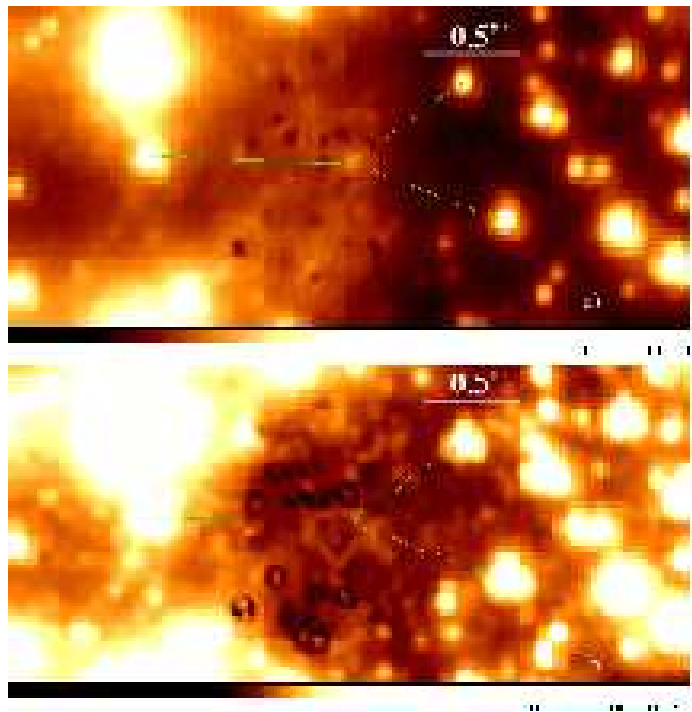}
\caption{\small
Results of the iterative star subtraction obtained for the 
30 August  (a) 23 September data (b) as described in 
Sect.~\ref{subsection:Iterative}.
}
\label{fig-3}    
\end{figure}

\begin{figure}
\centering
\includegraphics[width=9cm,angle=-00]{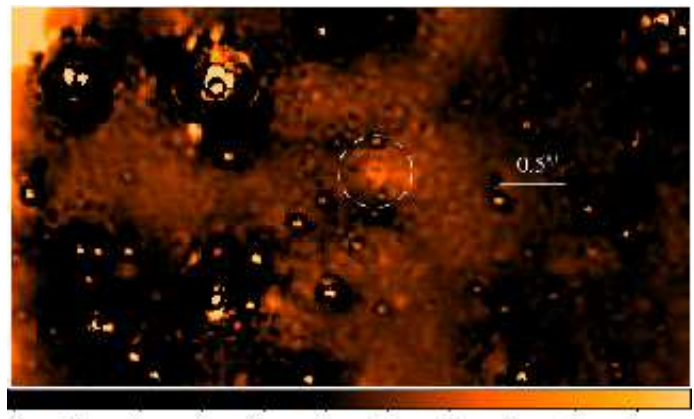}
\caption{\small
Result of the automatic PSF subtraction
obtained for the 23 September 2004 data.
}
\label{fig-4}    
\end{figure}

\begin{figure}
\centering
\includegraphics[width=8cm,angle=-00]{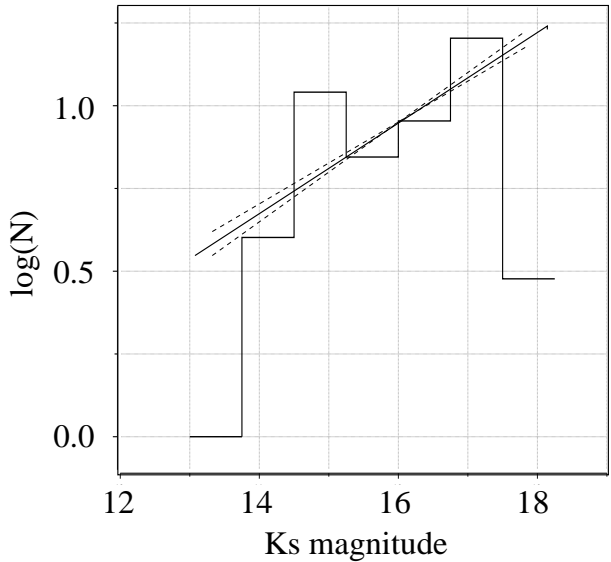}
\caption{\small
KLF histogram of the stars detected in the central field
derived from the 23 September 2004 data.
The straight full and dashed lines indicate the KLF slope of
0.21$\pm$0.02 found for radii of less than 6'' by Buchholz et al. (2009).
}
\label{fig-5}    
\end{figure}

We also analyzed the images with the StarFinder (Diolaiti et al. 2000) program package.
It not only provides accurate
point source photometry but also a reliable estimation of the diffuse
background emission. 
With a point spread
function (PSF) extracted from bright stars near
the guide star (in this case IRS 7), we performed 
photometry and astrometry on the deconvolved image via PSF fitting with 
the StarFinder package.
The resulting diffuse background emission is shown in Fig.~\ref{fig-4}.   
In addition to some residual emission from unresolved stars
this method also reveals extended emission centered on the position of SrgA*
within the dashed 0.5'' diameter circle centered on the position of SgrA*.
This finding agrees with the low-pass filtered image in Fig.~\ref{fig-2}.


\noindent


\noindent


\noindent

\noindent

\noindent
\noindent
\begin{figure}
\centering
\includegraphics[width=7.2cm,angle=-00]{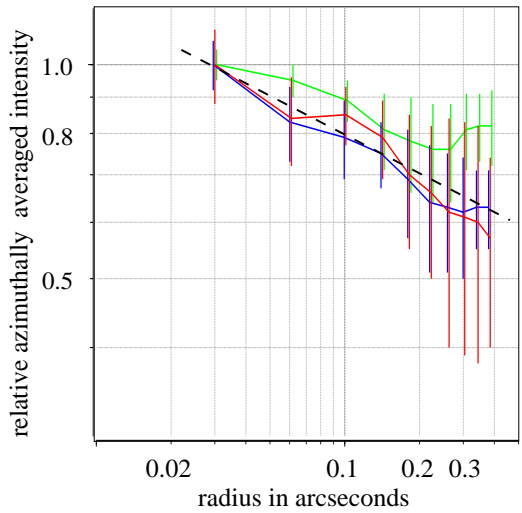}
\caption{\small
Azimuthal average of the diffuse background emission as derived from
the three different methods applied: low pass filtering (red), 
manual PSF subtraction (green), and automatic PSF subtraction (blue).
The curves 
(mean flux and 1$\sigma$ uncertainty per pixel) 
have been calculated in anuli of 39.8~mas (3 pixels) width 
and normalized to the value at 30~mas radius corresponding to one 
resolution element at 2.2$\mu$m.
The black dashed line marks a profile with an exponential 
decrease of 0.2.
}
\label{fig-6}    
\end{figure}

\noindent
\begin{figure}
\centering
\includegraphics[width=7.2cm,angle=-90]{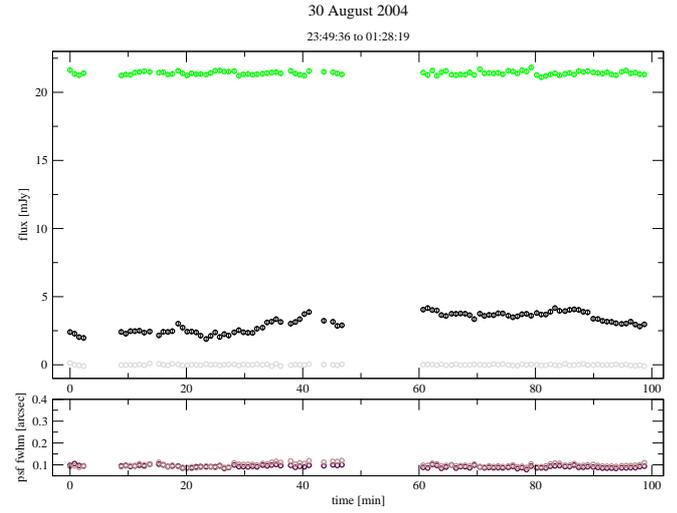}
\caption{\small
The 30 August 2004
light curve of the flux density measured in a 40~mas radius circular 
aperture centered on the position of SgrA* (black).
We show the light curve of the reference star S2 in green.
The bottom panel shows the FWHM seeing in arcseconds measured on the PSF 
extracted for the field.
The grey points show the background counts near SgrA*.
The width of the error bars is close to the size of the data points.
}
\label{fig-7}    
\end{figure}

\noindent
\begin{figure}
\centering
\includegraphics[width=7.2cm,angle=-90]{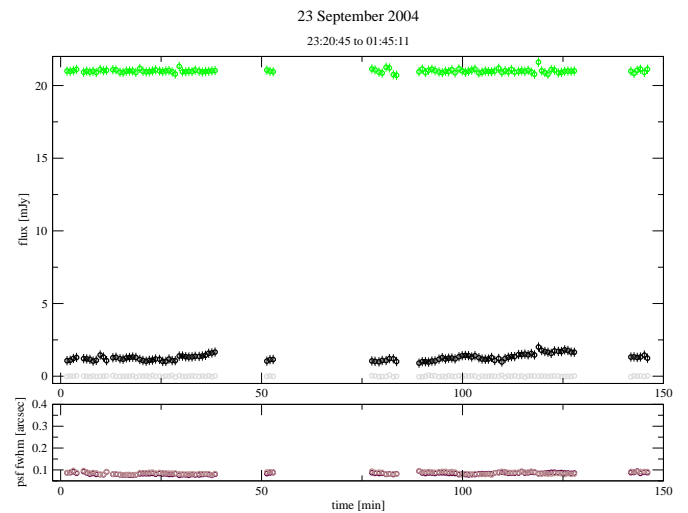}
\caption{\small
The same as Fig.\ref{fig-7} but for 23 September 2004.
}
\label{fig-8}    
\end{figure}

\section{Results and discussion}
\label{section:ResultsandDiscussion}

First we analyze the properties of the central cluster of high velocity stars
and compare them to the results of published investigations.
We can then confidently derive the flux density limits on SgrA*
and show that contributions to the lowest luminosity states from bremsstrahlung
are unlikely.
We then discuss the extreme luminosity states of SgrA*
in the framework of a synchrotron self-Compton mechanism.

\subsection{Imaging the S-star cluster and the detection of SgrA* in its low luminosity state}
\label{subsection:ImagingTheCusp}

The detected number of 51 stars within a radial distance of about 0.5'' 
from SgrA* results in a surface number density of 
68$\pm$8~arcsec$^{-2}$, taking the square-root as the uncertainty of that value.
This result agrees with the central number density of
60$\pm$10~arcsec$^{-2}$ given by Do et al. (2009b) in 
their Figs. 9 and 10.

Figure~\ref{fig-5} shows the KLF histogram (K-band luminosity function)
that can be derived for the stars detected in the central field.
The binning of 0.75 magnitudes allows for a sufficient number of bins
to clearly detect the linear slope in the double logarithmic plot and to
have a sufficiently large number of sources per bin (about 10) at the same time.
With d~log(N)/d~log(Ks)=0.3$\pm$0.1 the data compare well with the KLF slope of
0.21$\pm$0.02 found for the inner field ($R<6''$) by Buchholz et al. (2009).

Within the central 0.5~arcsecond radius region around SgrA* and in 
the magnitude interval ranging from Ks=16.75 to 17.50 no significant deviation
from a straight powerlaw line can be detected - implying that the
completeness is high and probably comparable to the $\sim$70\% value derived
for mag$_K$=17 by Sch\"odel et al. (2007).
In the interval ranging from Ks=17.50 to 18.25 the stellar counts drop
quickly to about 20\% of the value expected from the straight powerlaw line.

All three methods we used to correct for the flux density contribution of the stars in the
central 2'' reveal faint extended emission around SgrA* 
(see Figs.~\ref{fig-2}b, Fig.~\ref{fig-3}b and Fig.~\ref{fig-4}).
For the iterative and automatic PSF subtraction the PSFs were extracted 
with a radius of 1'', which is about twice the FWHM of the S-star cluster.
In case of a PSF misplacement a significant flux density contribution
to the central position can only come from the about five stars that are located within
one FWHM of SgrA*. They have a median brightness of about 2~mJy. To explain all of
the $\sim$2~mJy at the center by this effect, each star would have to 
provide about 0.4~mJy or 20\% of its flux density. This can only be realized by a 
systematic positional shift of these stars towards the center by about 1~pixel$\sim$13~mas
each. Still, the positional accuracy which is typically reached is on the order of a few tenths of
a pixel. Larger displacements result in a clearly identifiable characteristic plus/minus pattern 
in the residual flux distribution along the shift direction.
However, allowing for a maximum positional uncertainty of 1 pixel 
and approximating the independent shifts of five stars by five shifts of a single 
star that follows a random walk pattern, the expected effective displacement is 
1/$\sqrt 5$ pixels for a single shift of each star. It will result in a 5\% peak flux contribution of each 
of the 5 stars - assuming that the displacements are systematically towards the center, i.e. a total
maximum contribution of 0.4~mJy.
This implies that about one 20\%-30\% of the flux density detected at the center may be explained by a 
positional misplacement of the neighboring stars.
Therefore we assume that the bulk (more than 2/3) of this extended emission toward the center 
can be associated with the $\sim$0.5''-1'' diameter S-star cluster around SgrA* and is  
due to faint stars at or beyond the completeness limit reached in the KLF.

The extended residual emission is not distributed azimuthally uniform
about the position of SgrA*. This is not expected either. The shape
of the residual emission will always be dominated by the randomly
distributed few brightest stars that contribute to it.
Also the light of the central stellar cluster is not
distributed azimuthally uniform on any scale: On scales larger than
0.5~pc it is affected by dust absorption (e.g. extinction map by
Buchholz et al. 2009), on scales less than 0.5~pc the bright members of
the IRS16 cluster are predominantly located to the east, and during the past
decade in the central arcsecond
most of the brighter S-stars have been located to the south and east of SgrA*.

The diffuse background emission can be compared to the
projected distribution of stars $\Sigma(R)$$\propto$$R^{-\Gamma}$,
with $R$ being the projected radius.
Independent of the applied method we find that the azimuthally 
averaged residual diffuse background emission centered at 
the position of SgrA* decreases very gently as a function of radius
(Fig.~\ref{fig-6}).   
The ratio between the brightness measured within or at the edge of
the central resolution element (R$\sim$0.03'') 
and at a projected radius of 
e.g. R=0.2'' is 1.5$\pm$0.15,
corresponding to  $\Gamma_{diffuse}$=0.20$\pm$0.05.
This is consistent with the distribution of number density counts 
of the stellar populations in the central arcseconds
derived from imaging VLT and Keck data.
Buchholz et al. (2009) and Do et al. (2009b) 
find a $\Gamma$$\sim$1.5$\pm$0.2 for the young stars, but a much flatter distribution 
for the late-type (old) stars
with $\Gamma$$\sim$0.17$\pm$0.09 or even $\Gamma$$\sim$-0.12$\pm$0.09, respectively.  
For pure imaging Do et al. (2009b) quote
$\Gamma_{inner}$=0.19$\pm$0.06 inside a radius of R=3''.

In case of a relaxed stellar cluster around a supermassive black hole, stellar  
dynamics predicts to observe a cusp, i.e. a density increase toward 
the black hole with  $\Gamma$ = $\gamma-1$ = 0.5 to 0.75 (Bahcall \& Wolf 1976, 
Murphy 1991; Lightman \& Shapiro 1977; Alexander \& Hopman 2009). 
Here $\gamma$ denotes the exponent of the three dimensional 
distribution which does not suffer from projection effects 
but is harder to deduce from observed data.
Only in the case of very extreme stellar densities collisions  
can lead to a $\Gamma$ as low as 0.5. But the required high densities 
are not reached in the Galactic center. 

The observed values of $\Gamma$ are 
significantly lower than what is  predicted by theory. 
The principal reason is probably to be sought in the stellar population 
in the central  arcseconds, which towards Sgr~A* becomes 
increasingly dominated by young massive stars  that are too young to be 
dynamically  relaxed. 
The small value of the projected diffuse light 
exponent $\Gamma_{diffuse}$$\sim$$\Gamma_{inner}$
may therefore imply the absence of a pronounced, relaxed cusp of stars.

However, there may in fact exist a deficit of 
late-type stars near Sgr A* (see also  Genzel et al. 2003, Fig. 7). 
It may well be that most of the spectroscopically identified late 
type stars at small projected distances are at larger 
three dimensional radii $r$ $R$ from SgrA*.
In this case the observed $\Gamma_{diffuse}$$\sim$$\Gamma_{inner}$
will be a mixture between the much steeper $\Gamma$$_{early}$ 
for early type stars and the lower value $\Gamma$$_{late}$ for 
late type stars. A detailed discussion is given in 
Do et al. (2009b), Buchholz et al. (2009), Sch\"odel et al. (2007),  
and Genzel et al. (2003).

The main result for the present analysis of faint luminosity states of SgrA* is that the
observed flat background light and number density distribution 
described by the exponent $\Gamma_{diffuse}$$\sim$$\Gamma_{inner}$
as well as the high degree of completeness reached around Ks=17.5
allows us to clearly distinguish the emission of SgrA* against
the stellar light background at the center of the Milky Way.
This also shows that SgrA* is ideally suited to perform a case study
to investigate this super-massive black hole in its 
lowest activity state.

\subsection{Observing SgrA* at low luminosities}
\label{subsection:sgralow}

High angular resolution images in the near-infrared allow us to
unambiguously attribute bright dereddened flux levels to SgrA* 
(e.g. Genzel et al. 2003, Eckart et al. 2004, Ghez et al. 2004ab,
Hornstein et al. 2006, Do et al. 2009a).
The identification of fainter emission from
SgrA* becomes increasingly difficult, since one can expect 
confusion from 
the $\sim$1'' diameter S-star cluster that is centered on the 
location of Sgr A*.
Due to the high proper motions of the stars within the S-star cluster this
confusion also changes on the time scales of months to years.
From the location of SgrA* 
Do et al. (2009a) find 
observed $0.082\pm0.017$ mJy ($K^\prime = 17.2$) or dereddened $1.75\pm$0.36~mJy 
(assuming their value of $A_{K^\prime} = 3.3$)
as the the faintest observed $K^\prime$ flux density 
(see also Hornstein et al. 2002 and Eckart et al. 2006).
Using $A_{K^\prime}=2.8$ this results in a flux density of $1.1\pm$0.2~mJy

For the low flux density state Do et al. (2009a) reported red colors and variability 
compared to stellar sources. For the very low luminosity levels 
they report with an average  power law exponent of 
-0.17$\pm0.32$ a bluer spectrum than that reported earlier
by Hornstein et al. (2002) with a power law slope of $0.6\pm0.2$. 
The contamination from nearby (predominantly blue) stars is estimated to 
contribute a maximum of about 35\% 
of the flux to account for this difference in the spectral slope. 
Do et al. (2009a) conclude that even when the emission is 
faint a large fraction of the flux arising from the location 
of SgrA* is likely non-stellar and can be attributed 
to physical processes associated with the black hole. 

Our data allow us to provide a new independent estimate on the 
flux density of SgrA* in some of its lowest states  based on VLT data.

In a circular aperture of a diameter of
66~mas diameter (5 pixels or about one resolution element) we 
measure  dereddened flux densities
(using $A_K=2.8$)
1.5$\pm$0.4~mJy,
1.3$\pm$0.3~mJy, and
1.3$\pm$0.2~mJy for the 
low-pass filtering,
iterative, and 
automatic PSF subtraction method, respectively.

The uncertainties have been derived 
from the flux density measurements in the central 5$\times$5 pixel 
aperture and the upper limits of the measurement 
uncertainties per pixel as plotted in Fig.~\ref{fig-6}.
These upper limits are 
0.33~mJy,
0.24~mJy
and 0.17~mJy for the three methods, respectively.
Based on these data we derived an upper limit on 
the formal uncertainties in any random 5$\times$5 pixel aperture
in the central 0.6'' diameter field of
0.066~mJy,
0.048~mJy
and 0.034~mJy for the three methods, respectively.
We adopted these values for the central aperture.
In order to obtain a conservative estimate we added in 
each case these values five times,
so that the final uncertainty amounts to six times the value expected for 
an aperture of this size - based on the radial averages.
There is no clear detection of a source at that position.
If the observed flux density limit is fully attributed to 
SgrA* this corresponds to a clear non-detection of any point 
source at that position with a flux density limit of 2.4~mJy for SgrA* 
(K$_s$$\sim$16.3; observed reddened magnitude).
If all of it is stellar, the upper limit is defined by the uncertainty of the measurement
and the limit for a flux density from a non-stellar source at the position of SgrA* is 0.9~mJy 
(3$\sigma$, de-reddened and K$_s$$\sim$17.3 as an observed reddened magnitude).
While our data do not allow us to confirm the existance of a constant quiescent state,
our flux density limit is consistent with the flux given by Do et al. (2009a).
This flux density limit is also consistent with the light curves of SgrA* shown 
in Figs.\ref{fig-7} and \ref{fig-8}.
They were obtained in a 40~mas radius aperture centered at the position of SgrA*. 
The light curves were taken without removing any stars 
from the sourroundings and the aperture size of only 40~mas
was chosen to minimize the contamination
from neighboring stars.
The bottom parts of Figs.\ref{fig-7} and \ref{fig-8} show the FWHM of the PSF extracted on
a nearby star. These graphs show that the seeing and AO performance were very stable
and similar during the measurements on both dates.
Especially in the second half of the 30 August light curve in Fig.\ref{fig-7} SgrA* 
was sufficiently bright to be detected, while it was in a low state during the
entire time during which the 23 September light curve in Fig.\ref{fig-8} was taken.
Our flux density limit for the 23 September data is consistent with the completeness 
limit we reached in the central star counts 
(see Sect.~ \ref{subsection:ImagingTheCusp}).

\noindent
\begin{figure}
\centering
\includegraphics[width=9cm,angle=-00]{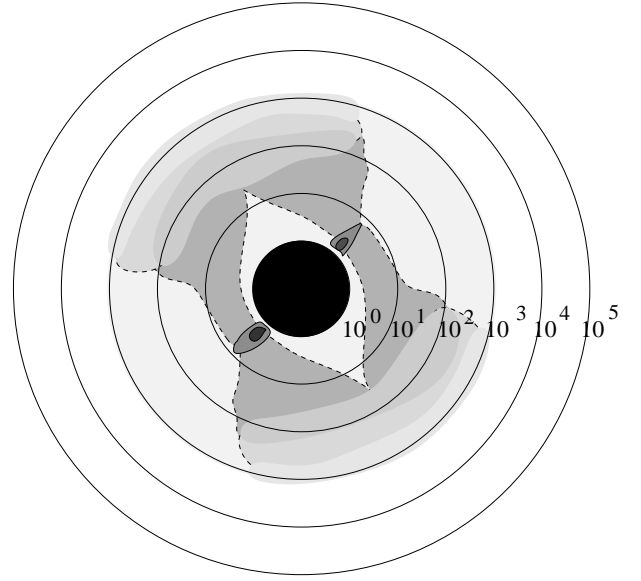}
\caption{\small
Sketch of the SgrA* surroundings within a 10$^5$ Schwarzschild radius.
The SMBH encloses the central 1~R$_s$ volume surrounded by an edge on 
seen disk with indicated spots.
The circles are logarithmically scaled and labeled 
with their radii in units of R$_s$.
During and after flares the spots expand 
to meet the accretion flow density of about 10$^7$cm$^{-3}$ at an
approximate radius of 1000~R$_s$.
The edge of the 1.4'' FWHM size 
Bondi sphere is located at a radius of about 
$\sim$8$\times$10$^4$R$_S$ for a 4$\times$10$^6$ \solm
(Baganoff et al. 2001, 2003). 
}
\label{fig-9}    
\end{figure}

\subsection{Bremsstrahlung emission from SgrA*}
\label{subsection:Brems}

\subsubsection{The quiescent state X-ray emission from SgrA*}
\label{subsubsection:X-rayLowState}

Baganoff et al. (2001, 2003) reported an extended (R$>$10$^3$ R$_s$)
thermal bremsstrahlung source co-spatial with the position of the
strongly variable point source at the location of SgrA*
(see also Quataert 2003).
The authors derive 
1.4''$\pm$0.14'' for the intrinsic FWHM size 
of the source. This corresponds to 
$\sim$8$\times$10$^4$R$_S$ for a 4$\times$10$^6$ \solm
black hole.
The scale of the extended structure centered on SgrA*
is consistent with the expected 
Bondi accretion radius (1''- 2''; Bondi 1952) for 
matter accreting hydrodynamically onto the SMBH.
For the bright flare emission of the X-ray point source 
bremsstrahlung is not a likely mechanism to produce
the observed NIR flux density excursions 
(Genzel et al. 2003, Yuan et al. 2003, Quataert, 2002).
In part motivated by the blue spectrum reported by Do et al. (2009a)
- assuming that not all of the blue character is due to 
stellar contamination -
we investigate here if the lowest NIR flux density values 
can be explained by a bremsstrahlung contribution from
a hot disk.

\begin{table*}[!htp]
\centering
\caption{Source component parameters for SSC models of the brightest X-ray flares from SgrA*.
\label{modeldata}}
{\begin{footnotesize}
\begin{tabular}{lrrclrrrrrrrrrr}
\hline
model & $\gamma_1$&$\gamma_2$&S$_{max,}$&$\alpha_{synch}$&R$_0$&$\nu_{max}$ & B   &S$_{MIR,}$ &S$_{NIR,}$  & S$_{NIR,}$ &S$_{NIR,}$&S$_{X-ray,}$\\
      &           &          &~~~$^{obs}$ &              &     &~$^{obs}$   &     &~$^{synch}$ &~$^{synch}$  &~$^{synch}$ &~$^{SSC}$ &~$^{SSC}$\\
label &           &          &[Jy]        &              &     & [GHz]      & [G] & [mJy]      & [mJy]      & [mJy]      &[mJy]     &[$\mu$Jy]    \\
      &           &          &            &              &     &            &     &  @11.8$\mu$m & @3.8$\mu$m & @2.2$\mu$m &@2.2$\mu$m&@2-10 keV \\
 \hline
$1~\sigma$ $\rightarrow$  & & &  $\pm$0.1&$\pm$0.1 &$\pm$0.1 &$\pm$250 &$\pm$10&$\pm$1.0&$\pm$1.0  &$\pm$1.0&$\pm$1.0 &$\pm$20 \\
 \hline
 A$\alpha$    &  360 & 2900 & 1.9 & 1.19 & 0.3 & 1850 & 18 &  59 & 15 & 8.0 & 0  & 0.65     \\
 A$\beta$     &  360 & 2900 & 3.2 & 1.25 & 0.3 & 1850 & 12 &  82 & 20 & 10  & 0  & 2.6     \\
 A$\gamma$    &  360 & 2900 & 1.3 & 1.05 & 0.2 & 1850 & 11 &  58 & 18 & 10  & 0  & 1.8     \\
       &      &      &     &      &     &      &    &     &    &     &    &  \\
 B$\alpha$    &  6   & 1300 & 2.8 & 1.10 & 0.3 & 2190 & 26 & 133 & 20 & 0.0 & 11 & 0.65     \\
 B$\beta$     &  6   & 1300 & 4.3 & 0.98 & 0.3 & 2190 & 18 & 226 & 26 & 0.0 & 15 & 2.6     \\
 B$\gamma$    &  6   & 1300 & 3.3 & 1.01 & 0.3 & 2190 & 20 & 219 & 17 &  0  & 10 & 1.8     \\
       &      &      &     &      &     &      &    &     &    &     &    &  \\
 $\xi$    &  360 & 2900 & 11.5 & 1.40 & 20.5 & 480 & 6 &  0 & 0 & 0 & 0 & $<$0.04 \\
       &      &      &     &      &     &      &    &     &    &     &    &  \\
 A$_{low}$    &  360 & 2900 & 0.5 & 1.10 & 0.1 & 2360 & 51 &  23 & 6.7 & 3.7 & 0  & 0.038     \\
 B$_{low}$       &  6   & 1300 & 0.8 & 1.30 & 0.2 & 1230 & 15 &  15 & 3.6 &  0  &3.7 & 0.037     \\
 Q$_{spot}$      &  70  & 2050 &0.08 & 1.30 & 0.1 & 1080 & 44 &  0.58   & 0.18    &0.1 & $<$0.01 & $<$0.010      \\
 \hline 
\end{tabular}
\end{footnotesize}}


\begin{center}
\caption{Model densities, lower SSC cutoff frequencies and lower limits on 
the source component masses
\label{modeldensity}}
{\begin{small}
\begin{tabular}{cccllccc} \hline
 model &   N$_0$ & c($\gamma_1^2 \nu_{max,obs}$)$^{-1}$& $\rho$ & mass &  \\
      & erg$^{-1}$cm$^{-3}$& $nm$          &  cm$^{-3}$  & M$_{\odot}$ &   \\
\hline
  A$\alpha$ & 0.25& 1.3 & 3$\times$10$^{7}$ & 3$\times$10$^{-16}$& \\
  A$\beta$  & 0.22& 1.3 & 6$\times$10$^{7}$ & 5$\times$10$^{-16}$& \\
  A$\gamma$ & 14  & 1.3 & 2$\times$10$^{8}$ & 5$\times$10$^{-16}$& \\
   &   &  &  & & \\
  B$\alpha$ & 0.74& 4100 & 2$\times$10$^{11}$ & 2$\times$10$^{-12}$ & \\
  B$\beta$  & 13  & 4100 & 3$\times$10$^{11}$ & 3$\times$10$^{-12}$ & \\
  B$\gamma$ & 7   & 4100 & 2$\times$10$^{11}$ & 2$\times$10$^{-12}$ & \\
   &   &    &  & & \\
  $\xi$ & 4$\times$10$^{-4}$ & 4.8    & 1$\times$10$^{6}$ & 3$\times$10$^{-12}$& \\
   &   &    &  & & \\
  A$_{low}$    &0.39 & 1.0  & 1$\times$10$^{7}$ & 3$\times$10$^{-18}$& \\
  B$_{low}$    &0.013& 7300 & 3$\times$10$^{11}$ & 8$\times$10$^{-13}$ & \\
  Q$_{spot}$   & 7$\times$10$^{-4}$ & 56    & 3$\times$10$^{7}$ & 9$\times$10$^{-18}$& \\
\hline
\end{tabular}
\end{small}}
\end{center}
\end{table*}


\subsubsection{Possible low state bremsstrahlung NIR luminosity?}
\label{subsubsection:BremsK-band}

We now estimate the possible contribution of the X-ray emitting gas that is associated 
with the compact disk or jet footpoint of SgrA* to its Ks-band luminosity.
The frequency and temperature dependency of the volume emissivity for the bremsstrahlung 
is
E$_{\nu}$d$\nu$ $\sim$ g($\nu$,T)$\times$exp(-$\frac{h \nu }{kT}$)d$\nu$.
Here the Gaunt factor is g($\nu$,T)$\sim$0.54$\times$ln(4.7$\times$10$^{10}$ T/$\nu$).
For the X-ray luminosity in the 2-10~keV range of a H/He plasma this
results in
$L(\frac{erg}{s}) \approx 10^{33} n_e^2 T_8^{1/2} R^3$ with
the source radius
$R$ given in $pc$, the electron density $n_e$ given in $cm^{-3}$, 
and the temperature $T_8$ given in unity of 10$^8$K.
The Gaunt factor 
implies that for a change in the observing frequency from
$\sim$10$^{18}$Hz in the X-rays to $\sim$10$^{14}$Hz in the NIR 
for constant T$>>$h$\nu$/k the Gaunt factor changes significantly
by a factor of ln(10$^{4}$)$\sim$10.
So in total the Ks-band luminosity expected 
over this frequency range 
is about 1000 times lower than the X-ray luminosity.
However, already for a $\sim$1~mJy Ks-band flux density contribution and high temperatures above
10$^{8}$K the predicted X-ray flux would well exceed the measured values.

Substantial flux density contributions due to bremsstrahlung can only occur if a
sufficient amount of 'cooler' gas (i.e. $<<$10$^8$~K) is mixed into the central accretion flow or disk.
Such a two phase medium has already been discussed in the context of flares from SgrA*
(Yuan et al. 2003, see also Page et al. 2004, Nayakshin \& Sunyaev 2003).
A possible quiescent phase at low infrared Ks-band flux density could in principle be provided 
through bremsstrahlung emission from a
cooler denser phase embedded in the hot accretion flow or in the outer parts of a potential accretion disk.
In luminous AGN a cool optically thick disk
with T$\sim$10$^5$K is believed to coexist with a hot optically thin corona with T$\sim$10$^9$K.
At the lower SMBH mass and luminosity of SgrA* such material 
could have an even lower temperature.
However, a lower limit to the temperature of such cool inclosure is given by the fact that no 
strong recombination line emission is observed towards SgrA*.
Only for temperatures above 10$^{6}$K the contribution to the total luminosity
from thermal bremsstrahlung is greater than that from recombination radiation.
This implies for 10$^{6}$K a Ks-band luminosity of
$L(\frac{erg}{s}) \sim  10^{29} n_e^2 R^3$.
At a distance of $\sim$8~kpc and a wavelength of 2.2$\mu$m
1~mJy of dereddened flux density corresponds to a luminostity
of $\nu$L$_{\nu}$=1.2$\times$10$^{34}$ ergs/s=3.1 \solar .

In order to obtain  1~mJy Ks-band flux density the product $n_e^2 R^3$ needs to be
on the order of 8$\times$10$^4$.
This allows us to estimate the required densities:
The Ks-band diffraction limit of the VLT UT4 corresponds to a projected linear size
of 2.4$\times$10$^{-3}$pc$\sim$8000~R$_S$ at the distance of SgrA*.
Similarly, 1~R$_S$ corresponds to 3$\times$10$^{-7}$pc.

For such an unresolved or even more compact source with sizes 
ranging from 1 to 8000~R$_S$ we find
electron densities from  
2$\times$10$^{12}$cm$^{-3}$ to
2$\times$10$^{6}$cm$^{-3}$ 
and Thompson scattering optical depths on the order of 1 to 0.01, respectively.
This implies that for an 'unresolved' bremsstrahlung emitting Ks-band source the
electron density is several orders of magnitude higher than the estimated 
density of 10$^7$cm$^{-3}$ of hotter gas 
in the accretion flow (Yuan et al. 2003).
Combined with the high optical depths this shows that a substantial 
bremsstrahlung contribution 
of a compact source to the Ks-band flux density 
even in the quiescent phase 
can therefore be regarded to be very unlikely
(see also Yuan et al. 2003).

\subsection{Synchrotron Self-Compton emission from SgrA*}
\label{subsection:SSC}

Rapid variability on time scales of a few minutes and the
strong indication of synchronous (polarized) NIR and X-ray
flux density variations linked with time delayed sub-mm/mm
flux excursions clearly suggests that a nonthermal emission
mechanism is at work. Synchrotron emission from compact source 
components that unavoidably scatter electromagnetic power into the
X-ray domain appear to be a prime choice for such a mechanism.
Below we investigate the highest and lowest luminosity states 
to validate the applicability of the SSC mechanism.

\subsubsection{Essentials of current SgrA* SED modeling}
\label{subsubsection:Essentials}

Generally the electron energy distribution $N(E)$ responsible for the 
overall SED of SgrA* is described by a power-law 
$N(E)=N_0E^{-p} = N_0 (mc^2 \gamma)^{-p}$ 
with a Lorentz factor $\gamma$ and an electron power-law index $p$.
This power-law exhibits several breaks at 

\begin{equation}
\gamma_{min} < \gamma_{cool} < \gamma_{max}~~~,
\end{equation}

\noindent
where $\gamma_{min}$, $\gamma_{max}$ and $\gamma_{cool}$ are
the minimum, maximum, and the 'cooling break' Lorentz factors, 
respectively
(Yuan et al. 2003, Quataert 2002, Mahadevan \& Quataert 1997).
Here $\gamma_{cool}$ marks the difference between the emission 
of a thermal distribution of electrons with
$\gamma_{min, thermal}$=$\gamma_{min}$
 and
$\gamma_{max, thermal}$$\sim$$\gamma_{cool}$
and the synchrotron emitting, relativistic electron distribution
with 
$\gamma_{min, synch}$$\sim$$\gamma_{cool}$
 and
$\gamma_{max, synch}$=$\gamma_{max}$.
The spectral number density ratio of both distributions
is assumed to be  
$\frac{N_{synch}}{N_{thermal}} \sim 1...5\%$
(e.g. Yuan et al. 2003).
For values at and below $\gamma_{cool}$ the heating of the thermal electrons through
synchrotron self-absorption is important.
For values above $\gamma_{cool}$ a more efficient
cooling through synchrotron radiation is important.

A further usually higher frequency break occurs due to a 
modification of the underlying relativistic electron distribution 
as a consequence of synchrotron cooling losses
(\"Ozel, Psaltis \& Narayan 2000, Kardashev 1962).
For an electron power-law index $p$ 
and for Lorentz factors larger than $\gamma_{min}$
the resulting optically thin electromagnetic spectrum
will follow a power-law distribution with a spectral index
$\alpha=(p-1)/2$.
In the case of no further injection of fresh electrons with 
the powerlaw index $p$ the spectrum will steepen at the frequency 
$\nu_{break}$ towards a spectral index of $(2p-1)/3$.
If fresh electron injection occurs the spectral index will 
only steepen towards $p/2$ and is therefore closer to the
original index of $\alpha=(p-1)/2$.

At this frequency  $\nu_{break}$ the synchrotron cooling time equals 
the time within which the relativistic electrons can escape the source.
In the case of SgrA* and at about 3.5~R$_{s}$ (approximate radius of the last 
stable orbit) it can be written as

\begin{equation}
\nu_{break} \sim 2.6 \frac{B}{30~G}10^{14}~~Hz
\end{equation}

\noindent
(Kardashev 1962, \"Ozel, Psaltis \& Narayan 2000, Yuan et al. 2003, Dodds-Eden et al. 2009).
The frequency $\nu_{break}$ is relevant if 
photons are emitted above that frequency, i.e.

\begin{equation}
\nu_{max} =2.8\times10^6 B \gamma^2 \ge \nu_{break}
\end{equation}

\noindent
(e.g. Marscher 1983).
For B$>$30$G$ and $\gamma_{max} \ge 1400$ both 
$\nu_{break}$ and $\nu_{max}$ lie just shortward of the NIR domain.
Here the values for  B and $\gamma_{max}$ result from model calculations that
allow us to describe the NIR and X-ray flare emission in the 
framework of an SSC model.
At frequencies above $\nu_{max}$ the inverse 
self-Compton emission dominates the overall spectrum.

In most cases modeling of the overall SED of SgrA* assumes that
the synchrotron quiescent emission (see Fig.1 in Yuan et al. 2003) 
is dominated by the relativistic tail of
the otherwise thermal Boltzmann distribution that describes the
quiescent (bulk) of the radio to sub-mm/FIR spectrum.
Flares are then attributed to a variation in the distribution of this
relativistic part of the accretion flow.
In addition the bulk of the quiescent X-ray spectrum is explained by
thermal bremsstrahlung due to the extended X-ray bright 
Bondi sphere surrounding SgrA*
(Baganoff et al. 2001, 2003, Liu et al. 2006, Yuan et al. 2003, Quataert 2002, Mahadevan \& Quataert 1997).

Here we consider the following additional points:
the flare spectrum of SgrA* can be explained independently of the 
steady, quiescent emission from the accretion flow. We consider 'fresh'
synchrotron components that become optically thick at THz 
frequencies and inverse Compton scatter into the X-ray domain.
We also investigate synchrotron,  bremsstrahlung, and thermal
emission from the foot point of the accretion flow or a 
temporary disk as possible significant contributors for the
unresolved source components seen in the NIR during the low luminosity states of SgrA*.

\subsubsection{Description and properties of the SSC model}
\label{subsubsection:SSCmodel}

We have employed a simple SSC model to describe the observed 
radio to X-ray properties
of SgrA* with the nomenclature given by
Gould (1979) and Marscher (1983).
Inverse Compton scattering models provide an explanation for
both the compact NIR and X-ray emission
by up-scattering sub-mm-wavelength photons into these spectral domains.
These models are considered
as a possibility in most of the recent modeling approaches
and may provide important insights into some fundamental
physical properties of the source.
The models do not explain the entire low frequency radio spectrum
and the bremsstrahlung X-ray emission that dominates the luminosity 
between X-ray flares.
They do successfully account for the flare events observed in recent
radio to X-ray campaigns though 
(Eckart et al. 2003, 2004, 2006ab, 2008ab, 2009,
Yusef-Zadeh et al. 2006ab, 2007, 2008, Marrone et al. 2009).

We assume a synchrotron source of an angular extent $\theta$. 
The source size can be as big as a few Schwarzschild radii.
The emitting source is assumed to become optically thick at a frequency
$\nu_m$ with a flux density $S_m$ and has an optically thin spectral
index $\alpha$ following the law $S_{\nu}$$\propto$$\nu^{-\alpha}$.
This allows us to calculate the magnetic field strength $B$ and the
inverse Compton scattered flux density $S_{SSC}$ as a function of the
X-ray photon energy $E_{keV}$. The synchrotron self-Compton spectrum
has the same spectral index as the synchrotron spectrum that is 
up-scattered 
i.e., $S_{SSC}$$\propto$$E_{keV}$$^{-\alpha}$, and is valid within the
limits $E_{min}$ and $E_{max}$ corresponding to the wavelengths
$\lambda_{max}$ and $\lambda_{min}$ (see Marscher et al. 1983 for
further details).

The emitting electrons follow a power-law distribution
$N(E)=N_0E^{-(2\alpha+1)}$
with energies $\gamma_1 mc^2<E<\gamma_2 mc^2$.
Here $\gamma_1$ and $\gamma_2$ are the electron 
Lorenz factors.
Maximum Lorentz factors 
for the emitting electrons of the order of 
typically 10$^3$ are required to produce a sufficient SSC flux in the
observed X-ray domain.

A possible relativistic bulk motion of the emitting source results 
in a Doppler
boosting factor $\delta$=$\Gamma$$^{-1}$(1-$\beta$cos$\phi$)$^{-1}$.
Here $\phi$ is the angle of the velocity vector to the line of sight,
$\beta$ the velocity v in units of the speed of light $c$, and
$\Gamma$=(1-$\beta$$^2$)$^{-1/2}$ Lorentz factor for the bulk motion.
Relativistic bulk motion 
is not needed to produce sufficient SSC flux density, but 
we have used modest values for 
$\Gamma$=1.2-2 and $\delta$ ranging between 1.3 and 2.0 (i.e. angles $\phi$ 
between $10^{\circ}$ and $45^{\circ}$)
since they will occur
in cases of relativistically orbiting gas as well as relativistic 
outflows - both of which are likely to be relevant to SgrA*.

The electron spectral number density in units of $cm^{-3}keV^{-1}$ is given by
\begin{equation}
N_0=\kappa(\alpha,d)\theta^{-(4\alpha+7)}\nu_m^{-(4\alpha+5)}S_m^{2\alpha+3}\delta^{-(2\alpha+4)}~~.
\end{equation}

\noindent
Here S$_m$ and and $\nu_m$ are given in units of Janskys and GHz, respectively.
For $\kappa(\alpha,d)$ see Marscher (1983).
With $E=\gamma mc^2$
this results in a total number density in $cm^{-3}$ of electrons that
participate in the SSC emission process of
\begin{equation}
\rho=mc^2 N_0 \int^{\gamma_2}_{\gamma_1} N(\gamma) d \gamma =
N_0 \frac{(mc^2)^{-2\alpha}}{2\alpha}
 (\gamma_1^{-2\alpha} - \gamma_2^{-2\alpha})~~.
\end{equation}

\noindent
This quantity can be used to validate the emission mechanism 
in comparison to other number density estimates obtained
for different source regions.

The radio polarization data of SgrA* (Bower et al. 2004, Marrone et al. 2007) 
indicate that of the total available stellar mass loss of 
10$^{-3}$\solm yr$^{-1}$
only a small fraction of
10$^{-7}$\solm yr$^{-1}$
is close to the central black hole and available for accretion.
This applies a mean daily accretion of 3$\times$10$^{-10}$\solm.
Assuming a flare lasts for 100 minutes this corresponds to 
an accretion mass load of typically 2$\times$10$^{-11}$\solm per flare.
The SSC models in Table~\ref{modeldensity}
result in estimates that 
lie one to five orders of magnitude below this upper mass limit per flare.
Here models A$\alpha$ to A$\gamma$, in which the NIR fluxes are produced via optically 
thin synchrotron radiation, are favored since the required masses lie well below 
the available accretion mass load (as it is for the quiescent models).

\subsubsection{SSC model of the bright flares of SgrA*}
\label{subsubsection:SSCBright}

For X-ray flares of up to several times the quiescent emission the SSC models 
with source components that become optically thick in the few THz domain
provide a successful description of the compact flare emission that  
originates from the immediate vicinity of the central black hole
(e.g. Eckart et al. 2009 and references therein, Marrone et al. 2009; see also 
Fig~\ref{fig-9}).   
This description also allows the explanation of the correlation between
NIR/X-ray flares and time delayed radio flares
(e.g. Eckart et al.  2008ab, 2009,
Yusef-Zadeh et al. 2008, Marrone et al. 2009).
In order to validate the applicability of this emission mechanism 
it is useful to study the bright flare emission associated with SgrA*.

In Table~\ref{strongflares} we list important properties 
of the three brightest X-ray flares that have been observed to date
(Baganoff et al. 2001, Porquet et al. 2003, 2008).
The NIR flare emission shows observed spectral indices of
$\alpha$$\sim$0.6 (Ghez et al. 2005ab, Hornstein et al. 2006) or even
steeper (Eisenhauer et al. 2005, Gillessen et al. 2006).
If the X-ray flare flux density is due to the SSC spectrum of the synchrotron component that
gives rise to emission in the NIR then the spectral indices should be the same.
While a spectral index of $\sim$0.6 is consistent with the Chandra value,
the XMM flares require a steeper spectral index.
The spectral index averaged over several flares observed with Chandra results in 
$\alpha$=$\Gamma$-1=0.3$\pm$0.5 (Baganoff et al. 2001).
This is consistent with both the XMM data and the 
index of  $\alpha$$\sim$0.6 reported by Ghez et al. (2005ab) and  Hornstein et al. (2006).
\\
\\
{\it General modeling results:}
Our modeling results show that synchrotron flare components that become 
optically thick in the few THz domain stay optically thin throughout the 
MIR/NIR domain, and self-Compton scatter into the X-ray domain can 
fully account for the observed properties of the brightest flares from SgrA*.
We use maximum electron Lorenz factors $\gamma_2$$\sim$10$^3$ resulting in the
fact that the X-ray emission is always inverse self-Compton scattered.
This delivers synchrotron radiation up to near-infrared wavelengths and results
in the fact that the X-ray emission is always inverse self-Compton scattered.
Source component parameters for SSC models of the brightest 
X-ray flares from SgrA*
are given in Tables~\ref{modeldata} and~\ref{modeldensity}.

\begin{figure}
\centering
\includegraphics[width=9cm,angle=-00]{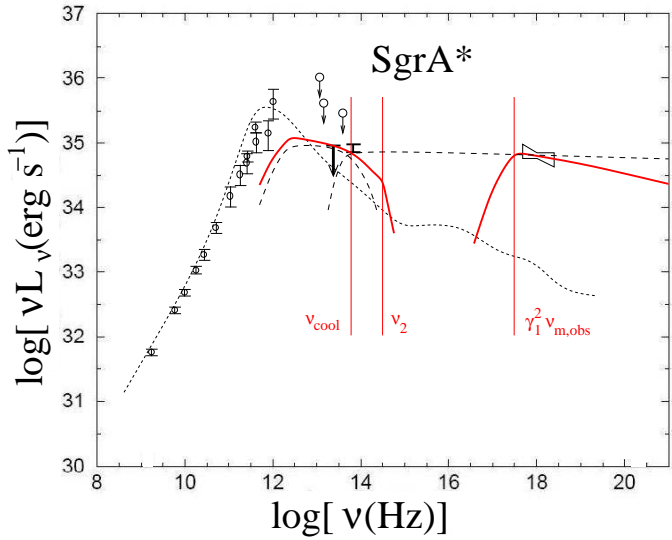}
\caption{\small
Two SSC models A (red) and B (black dashed)
describing the April 4 2007, flare data (see Table~\ref{strongflares},
Tables~\ref{modeldata} and~\ref{modeldensity}).
}
\label{fig-10}    
\end{figure}

Here model labels (col.~1) $\alpha$, $\beta$ and $\gamma$ indicate the
bright flares listed in Table~\ref{strongflares}.
Model labels A and B indicate models in which the the MIR/NIR flux
densities are accounted for in different ways.
In particular
in models A$\alpha$-$\gamma$
the observed peak MIR/NIR flux densities are accounted for by the 
high frequency end of the synchrotron spectrum.
In models B$\alpha$-$\gamma$ they are accounted for 
by the low frequency end of the inverse self-Compton scattered spectrum.
Among themselves the A models (straight red line in Fig.\ref{fig-10})
and the B models (black long dashed line in Fig.\ref{fig-10})
look very similar for the different flares labeled with
$\alpha$, $\beta$ and $\gamma$ in Fig.\ref{strongflares}.
Within the uncertainties infrared and X-ray flux densities as well as
the X-ray spectral indices are met by the
required model spectral indices. As an example (and since it this model is
discussed in more detail) we show 
the synchrotron flare spectra and the 
corresponding SSC spectra in Fig.\ref{fig-10}
for the models A$\alpha$ and B$\alpha$.

A global variation of a single parameter
by the value listed in the first row and the corresponding column of
Table~\ref{modeldata} results in an increase of $\Delta \chi = 1$.
Here global variation  means:
adding the 1$\sigma$ uncertainty
for a single model parameter but for all source components
in a way that a maximum positive or negative
flux density deviation is reached.
Models with significantly different component fluxes, sizes, and
cut-off frequencies  fail to match the observed fluxes by more than 30\%.
A more detailed description of the modeling procedure is
given in Eckart et al. (2009).
Models A$_{low}$ and B$_{low}$
give the source component for a single and multiple spot 
(multiples of the dominant spot described by model Q$_{spot}$)
quiescent state model.

There are two approaches to model the NIR/X-ray spectra:
1) The observed peak MIR/NIR flux densities can be accounted for either by the 
high frequency end and of the synchrotron spectrum (models A$\alpha$-$\gamma$, in 
Tables~\ref{modeldata} and~\ref{modeldensity}) 
or 2) by the low frequency end of the inverse self-Compton 
scattered spectrum (models B$\alpha$-$\gamma$).
Here the upper cutoff frequency of the scatter spectrum $\gamma_1^2 \nu_{max,obs}$
has a special importance.
With $\nu_{max,obs} \sim 2.2~THz$ and $\gamma_1 \sim 6$
we find that this cutoff lies in the few micrometer MIR/NIR domain.
With $\gamma_1 \sim 1300$ it lies at about 1keV just below the lower 
energy cutoff of the X-ray Chandra and XMM observatories.
In general the models A$\alpha$ to A$\gamma$ are constrained by the 
synchrotron spectral index $\alpha_{synch}$ 
defined by the  THz and NIR/MIR flux densities.
Models B$\alpha$-$\gamma$ are constrained by the 
SSC spectral index (that equals  $\alpha_{synch}$)
defined by the  NIR/MIR, the X-ray flux densities and spectral index.
Therefore 
$\alpha_{synch}$ cannot be steeper than about unity
in these SSC models.
\\
\\
{\it Possible peculiarities in the relativistic electron spectrum:}
A modification in the upper cutoff of the relativistic electron spectrum
(e.g. Lorentz factor $\gamma_2$) 
may also have a strong influence on the X-ray output of the emitting source component.
An increase in $\gamma_2$ may reflect a more efficient acceleration mechanism.
Synchrotron losses will lead to a 
decrease in $\gamma_2$ or to the involvement of a cooling break.
The width of the relativistic electron distribution for flare models
A$\alpha$ to A$\gamma$ is narrower than that of models B$\alpha$ to B$\gamma$ or even
models which explain additionally the low frequency radio part of the SgrA* SED
(Yuan et al. 2003, Quataert 2002, Mahadevan \& Quataert 1997).
Kardashev (1962) points out that the higher frequency synchrotron 
cooling break does not become displaced towards 
lower frequencies with time if the energy distribution of the injected 
relativistic electron spectrum has a very small width.
A narrow relativistic electron spectrum may therefore support a less 
variable cooling break frequency during the main part of the flare.

For models A the lower energy cutoff $\gamma_1$ lies above 
the typical values for $\gamma_{cool}$ at which efficient
cooling through synchrotron radiation sets in.
This suggests that the electron spectra responsible for the 
brightest SgrA* flares may have a different origin from those in the 
lower energy Boltzmann distribution which explain the radio spectrum.
\\
\\
{\it The April 4, 2007, flare:}
The flare event of April 4, 2007 (see Table~\ref{strongflares}), has also 
been simultaneously observed at the 11.8$\mu$m and 3.8$\mu$m wavelength.
Dodds-Eden et al. (2009) discuss several emission mechanisms.
They rejected a pure inverse Compton model because of too high magnetic fields
and a pure power-law model due to a mismatch of the MIR/NIR flux densities and 
spectral index information.
They strongly disfavor their SSC model in which the 
flare component becomes optically thick in the 
NIR with an exceptionally high magnetic field strength
(see the comment in Sect.~4.1.3 of Eckart et al. 2009). 
Their favorite model for the flare consists of a power-law spectrum in combination 
with a cooling break in the NIR/optical wavelength domain.
They do not discuss the unavoidable self-Compton contribution to the emission
that must originate from the bulk of the photons at the self-absorption 
turnover scattered by the relativistic electrons.
The lower cut-off frequency of this power-law solution 
remains unspecified as well. 
Previous strong X-ray or NIR flare events have been shown to be 
linked to delayed radio flares
(Eckart et al. 2006a, 2008b, 2009, Yusef-Zadeh et al. 2007, 2008, 2009,
Marrone et al. 2009). 
So the relation of the power-law flare solution to the mm/radio regime
remains unclear.

\noindent

\noindent
\begin{figure}
\centering
\includegraphics[width=8cm,angle=-00]{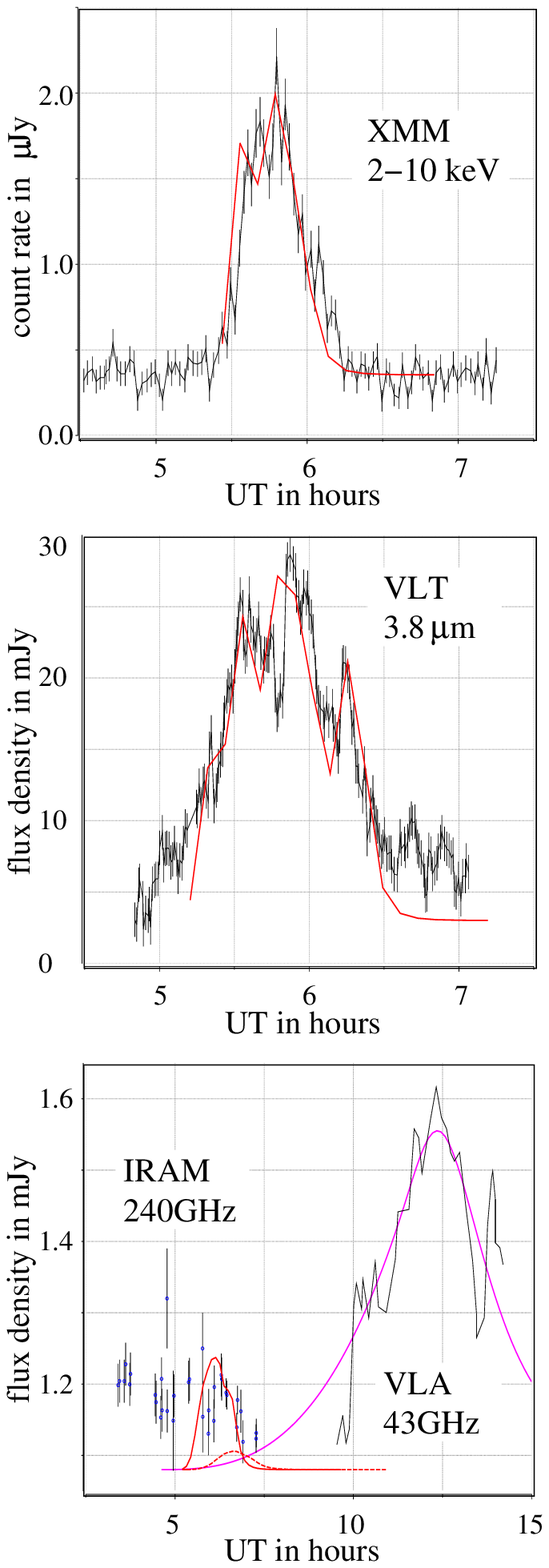}
\caption{\small
A combined SSC and adiabatic expansion model compared to the
SgrA* flare event on April 4, 2007.
The data are described in 
Porquet et al. (2003), Dodds-Eden et al. (2009), Yusef-Zadeh et al. (2009). 
The 240~GHz and 43~GHz response of the NIR/X-ray flare is plotted 
as a  solid and dashed red line, respecitvely. The probably unrelated 43~GHz response
of the flare component $\xi$ with the parameters listed in Table~\ref{modeldata}
is plotted as a solid magenta line.
}
\label{fig-11}    
\end{figure}

The SSC model that we present here for the the April 4, 2007, flare 
is not covered by Dodds-Eden et al. (2009).
Here we cover it in more detail and take it as an example for the bright 
luminosity state of SgrA*.
The results may be used to explain the 
properties of mm/sub-mm flux density variations that have been observed in 
several other cases
(e.g. Eckart et al.  2008ab, 2009,
Yusef-Zadeh et al. 2008, Marrone et al. 2009).

For the April 4, 2007 flare event
our SSC model A$\gamma$ agrees with the simultaneous 3.8$\mu$m flux density, the 3$\sigma$ 
flux limit at 11.8$\mu$m and also the
spectral index in the X-ray domain corrected for hydrogen absorption.
A spectral decomposition is given in Fig.~\ref{fig-10} and a comparison of the 
light curves with model A$\gamma$ and $\xi$ (see below) is displayed in Fig.~\ref{fig-11}.
For the favored model A in Fig.~\ref{fig-10} the relevant cutoff frequencies are 
labeled and marked with a vertical thin solid line.
For model B these frequencies are all close to 7$\times$10$^{13}$~Hz.
The model data are compared to the SED of SgrA* in its quiet state.
The radio to sub-mm measurements (Markoff et al. 2001; Zhao et al. 2003)
are time averaged measurements, and the error bars include variable emission
of up to 50\%. As open black circles we show 30$\mu$m, 24.5$\mu$m and 8.6$\mu$m 
upper limits taken from Melia \& Falcke (2001).
The X-ray information is taken from Porquet et al. (2008).
The short dashed line shows the quiescent model by Yuan et al. (2003).

With a magnetic field of 10~G the optically thin synchrotron spectrum extends up to 
$\nu_{break}$=c (3.4$\mu$m)$^{-1}$ with a spectral index of $\alpha_{synch}$=1.05. 
Above this the spectrum steepens until  the upper synchrotron cutoff
$\nu_{max}$=c (0.85$\mu$m)$^{-1}$. 
For a spectral index of unity, i.e. p$\sim$3, and 
with the assumption of a fresh injection of relativistic electroms the spectrum will 
steepen to a value of $p/2$$\sim$1.5. This value fully agrees with the 
spectral index between the infrared H- and Ks-band of $\alpha_{HK}$=1.4$\pm$0.2 
which was used as an additional fitting criterion by Dodds-Eden et al. (2009).

The model can be further validated by comparing the involved electron spectral number density
with values inferred from other data or different models.
The value of 14$cm^{-3}keV^{-1}$ of the electron spectral number density 
results in number densities for the relativistic electrons of 2$\times$10$^{8}$$cm^{-3}$.
Part of the emitting particles may be due to electron/positron pairs.
Integrating over the SSC source component volume of 0.2~$R_0$ diameter we
obtain 5$\times$10$^{-16}$~\solm as a lower limit to the total mass of this source 
component within a compact jet base that feeds an extended 
overall jet or outflow with a low surface brightness
(Falcke \& Markoff 2000, 2001, Markoff 2005, Markoff et al. 2007).
If a structure of the size and density described above expands 
at highly relativistic velocities,
it can reach the flow density of 10$^{7}$cm$^{-3}$ (Yuan et al. 2003).
This can happen on an orbital time scale
of a spot which is close to the last stable orbit
and on less than one flare time scale
(see Fig.~\ref{fig-9}).   
 
For the SSC model B$\gamma$ 
explains the NIR flux densities shortward of about 4$\mu$m through
the self-Compton scattered spectrum the 
electron spectral number density and source component mass. 
This mass is four orders of magnitudes
higher than for model A$\gamma$.
Reaching the flow density of 10$^{7}$cm$^{-3}$ in one flare time scale
and matching the 
total estimated  accretion rate onto SgrA* is possible as well.
Still, the predicted 11.8$\mu$m flux density is well above the upper limit
obtained for the April 4, 2007 flare. Therefore models in which the Ks- and L-band 
flux densities are produced via the self-Compton scattered spectrum are 
not favored.
\\
\\
{\it Modeling the April 4, 2007 NIR/X-ray light curves:}
Dodds-Eden et al. (2009)
report that the width of the X-ray flare is significantly smaller 
than the on of the corresponding overall L-band flare.
This implies that the efficiency of producing X-ray emission was lower 
before and after the main part of the flare.
In the framework of the evolving spot model presented in Eckart et al. (2008a)
this may be explained by an evolution of the spot characteristics
similar to the models presented by Hawley \& Balbus (1991, 1998) and Yuan et al. (2008).
The recent theoretical approach of hot spot evolution due to shearing is
highlighted in Eckart et al. (2008a) and Zamaninasab et al. 
(2008, 2009; see also Pech\'a\v{c}ek et al., 2008).
Yuan et al. (2008) explicitly solved the relativistic hydrodynamics and 
found initial expansion velocities of less than 0.01c close to
the accretion disk followed by relativistic expansion towards the end of the flare.
A compression of a single spot or merging of two spots 
due to differential rotation within a viscous disk may explain the 
different widths of the flares in the X-ray and L-band light curve.
A variation of the source size $\theta$ 
or the peak flux density $S_{\nu}$ of the scattered synchrotron component 
by only 20\% to 30\% will result in the 
required significant increase or decrease in the SSC scattering efficiency, i.e. the scattered
SSC X-ray flux density $S_{SSC,X-ray}$$\propto$$S_m^{2(\alpha+2)}$$\theta^{-2(2\alpha+3)}$.
Similarly other differences between the NIR and X-ray light curves may be due to different 
or even varying flux density contributions and scattering efficiencies 
from a source component responsible for an underlying main flare and components 
responsible for the sub-structures.
\\
\\
{\it Modeling the April 4, 2007 radio light curves:}
Yusef-Zadeh et al. (2009) report radio measurements that were 
carried out at 240~GHz (30m IRAM) during and at 43~GHz (VLA) 
following the April four NIR/X-ray flare.
The 240~GHz data cover the entire NIR/X-ray flare event at a
flux density level of about 1.2~Jy with no significant flux density variation.
After a ('transatlantic') gap in the data a strong increase and decrease
of the 43~GHz flux density with a peak of about 0.4~Jy is 
detected. The entire 43~GHz flare has a full bottom width
(at a level of about 1.1~Jy) of about 4 hours 
and peaks at 43~GHz about 6.5 hours  past the NIR/X-ray peak.

Performing an adiabatic expansion calculation (van der Laan, 1966)
based on models A$\gamma$ and B$\gamma$
using an expansion speed of 0.007c
places the 43~GHz afterglow of the NIR/X-ray flare close to 
or even into
the gap between the 100~GHz and 43~GHz measurements.
We use 0.007c as a typical value close to the expansion speeds found for 
other SgrA* flare events for which adiabatic expansion models have been 
applied (see e.g. Eckart et al. 2008b, 2009, Yusef-Zadeh et al. 2008).  
As an example of such a calculation we show in Fig.~\ref{fig-10} 
a combined SSC and adiabatic expansion model
based on the data for A$\gamma$ given in Table~\ref{modeldata}.
Here we assume that the 240~GHz and 43~GHz flux density level 
above which we have to consider the excess flare flux is at 
1.08~Jy - which is at the minimum flux density level 
given by the data for this event.
The predicted 240~GHz and 43~GHz peak brightnesses 
lie well within the noise of the IRAM measurements and the
following 43~GHz data.

A lower expansion velocity like 0.002c gives the time delay needed
to meet the bight 43~GHz flare peaking at about 12:30~UT
but still fails to reproduce its high radio flux.
Since measurements around this time have not given any hint for a possible
NIR or X-ray counterpart for this event, it can only be explained 
by a low frequency flare that originated from a synchrotron component
with low turnover frequency and no significant predicted NIR or X-ray flux density
(like component $\xi$ given in Table~\ref{modeldata}).
We therefore regard component $\xi$ as unrelated to the main NIR/X-ray flare
which happened six hour earlier.

The SSC models listed in Table~\ref{modeldata} have small source component sizes
to provide a sufficient inverse Compton scattering efficiency to 
explain the large observed X-ray flux densities.
This results in comparatively narrow light curves of the individual components
and requires the observed broader flare profile
to be modeled with six source components. 
This may explain the presence of substructure in the NIR and X-ray light curves.
We place the components 1 through 6 at 5:24, 5:37, 5:51, 6:00, 6:15, and 6:21~UT
with the flux ratios between them of 0.6:1:1:0.8:0.5:0.4. 
We also allowed components 1, 5, and 6 to be larger 
by $\sim20\%$ compared to the central components.
An increase in size or a decrease in flux density lowers the scattering 
efficiency (see above) in a way that the weaker components 1, 5, and 6 do not 
contribute significantly to the X-ray flare.
With a spectral index of 1.05 and the dependency in scattering efficiency given 
above this results in an X-ray flux density ratio 
between the components of 0.01:1:1:0.3:0.003:0.001.
This may explain the different widths of the flares in X-ray and L-band light curve
and allows us to model the broader NIR flare profile.

Model B$\gamma$ in Table~\ref{modeldata} is not well suited to explain the wings of the
NIR light curve in comparison to the X-ray flare profile since the flux densities 
in both wavelength domains are described by the same SSC scattered spectrum.
Involving steeper SSC spectra for the wings violates the 11.8$\mu$m flux 
density limit.
Therefore the 4 April, 2007, flare event provides additional support
for the assumption that - at least in this case - the NIR luminosity is due to 
optically thin synchrotron emission rather than SSC scattered flux.

\subsubsection{SSC modeling of a weak X-ray flare}
\label{subsubsection:SSCWeak}

In Tables~\ref{modeldata} and \ref{modeldensity} we also list modeling 
results for the so far weakest statistically significant
X-ray flare that was modeled (Eckart et al. 2002). This flare was also the first that
was detected simultaneously in both the NIR and X-ray domain.
The SSC model A$_{low}$ which explains the NIR flux densities through the high frequency end of 
the associated synchrotron spectrum describes the flare successfully. 

SSC models (B in Tables~\ref{modeldata} and \ref{modeldensity})
that explain the NIR fluxes through the inverse Compton scattered part of the spectrum 
are strongly constrained by the NIR/X-ray spectral index determined by $\alpha_{synch} \sim 1.0$.
While they do not result in a satisfactory fit of the observed 
data for the very bright X-ray flares 
(see Sect.~\ref{subsubsection:SSCBright}),
the situation for weak X-ray flares is different.
Here $\alpha_{synch}$$\sim$1.3 and the corresponding 
electron spectral number density and source component masses are of the same order as for the
SSC models A$\alpha-\gamma$.
Therefore both model approaches, A$_{low}$ and B$_{low}$, result in acceptable explanations 
for the NIR and X-ray fluxes and the spectral indices during weak flares.

\subsubsection{SSC modeling of the lowest states of SgrA*}
\label{subsubsection:SSCLow}

Eckart et al. (2006) have modeled the 
emission from SgrA* in its low flux density state.
They find that the upper limits of the compact X-ray emission in the 'interim-quiescent' (IQ), 
low-level luminosity states of SgrA* are consistent with
a SSC model that allows for substantial contributions from both the 
SSC and the synchrotron part of the modeled spectrum.
In these models the X-ray emission of the point source is well below 
20$-$30~nJy and contributes considerably less than half of the X-ray flux density 
during the weak flare event reported by Eckart et al. (2004).
The flux densities at a wavelength of 2.2$\mu$m are of the order
of 1 to 3~mJy
(see Table~\ref{weakflares} and Figs.~\ref{fig-7} and \ref{fig-8}). 
Eckart et al. (2006) list representative SSC models for the low flux state.
For their models IQ1-IQ3 
the source component has a size on the order of 0.2 to 2 Schwarzschild radii
with an optically thin radio/sub-mm spectral index ranging from
$\alpha$$\sim$1.0 to 1.3, a value similar to the
observed spectral index between the NIR and X-ray domain.

\begin{figure}
\centering
\includegraphics[width=9cm,angle=-00]{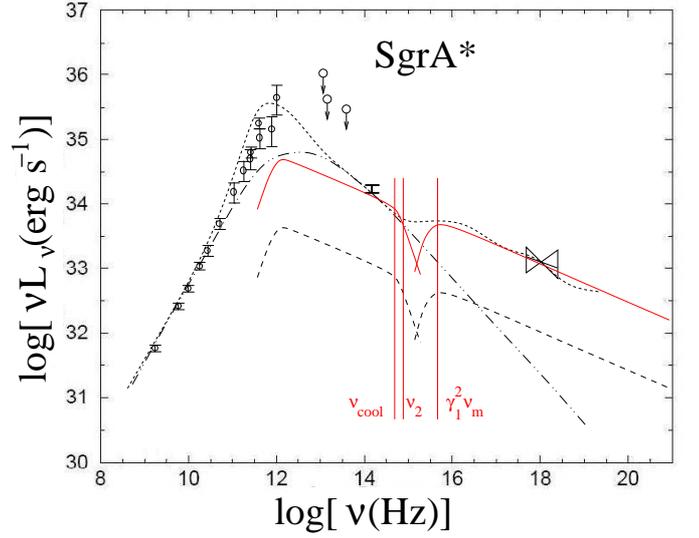}
\caption{\small
A single spot SSC quiescent spectrum (black dashed)  
and a corresponding multi-spot SSC quiescent spectrum. 
}
\label{fig-12}    
\end{figure}

In Tables~\ref{modeldata} and~\ref{modeldensity} we give the detailed 
model data for a single spot in which the 
MIR/NIR flux densities are accounted for by the 
high frequency end of the synchrotron spectrum (A$_{low}$) or
by the low frequency end of the inverse self-Compton scattered spectrum (B$_{low}$).
These models account for the quiescent source spectrum given by Yuan et al. (2003),
as shown in Fig.~\ref{fig-12}.  
For  model B$_{low}$ the electron spectral number density value
and total masses integrated over the SSC source component volume 
are large for a quiescent state compared to the flaring state 
($\sim$3$\times$10$^{11}$$cm^{-3}$ and 8$\times$10$^{-13}$~\solm)
and to the assumed accretion flow density of 10$^{7}$cm$^{-3}$.
Therefore model A$_{low}$ is favored.

\noindent

As an alternative to a larger continuous disk component one may assume a 
continuous disk or a disk consisting of several spots with dominant source 
components that contribute only about 0.1~mJy (model Q$_{spot}$)
to the Ks-band flux density each and have 
electron spectral number densities on the order of
$\sim$10$^{7}$$cm^{-3}$ 
and total masses integrated over small SSC source component sizes
of 0.1 R$_s$ with a diameter of 
a few times 10$^{-19}$~\solm.
Here the magnetic field strengths are comparable to those of the larger flare
components, and the densities are comparable to the 
assumed accretion flow density of 10$^{7}$cm$^{-3}$. 
In Fig.~\ref{fig-12} - as an example -
20 spots as described by model Q$_{spot}$ in 
Tables~\ref{modeldata} and~\ref{modeldensity} produce a significant part
of the quiescent spectrum of SgrA* in the Ks-band, as one would 
expect for a low level quiescent state of SgrA*.
The 20 spots we used in Fig.~\ref{fig-12} have different relativistic 
boosting factors as they would occur on an inclined circular orbit of a few
Schwarzschild radii around the SgrA* SMBH
(see Tables~\ref{modeldata} and~\ref{modeldensity},
and data in Eckart et al. 2004). 
In Fig.~\ref{fig-12} relevant cutoff frequencies are 
labeled and marked with a vertical thin solid line.
The quiescent X-ray information is taken from Baganoff et al. (2001, 2003).
The short dashed line shows the quiescent model by Yuan et al. (2003).
The double-dot-dashed line shows the contribution of the relativistic electrons to this 
quiescent model.

Therefore both a larger continuous synchrotron emitting disk or a 
faint disk spotted with SSC components are attractive 
models that may explain a possible low state of SgrA*. 
Any quiescent state flux density (e.g. measured in the NIR) 
would then be expected to be variable and polarized.
High angular resolution and sensitivity - as it will be provided by the
future large telescopes (ELTs) - are required to measure the quiescent state
flux density contribution with sufficient precision.

\subsection{Thin disk NIR luminosity}
\label{subsection:ThinDisk}

As another way to explain the Ks-band flux density especially in an assumed
quiescent state of SgrA*, we explore the possible presence of a small 
permanent disk that radiates as a black body. 
Eckart et al. (2006) show that at least for the flaring state of SgrA* a 
disk may exist. For an optically thick disk seen at an inclination angle $i$ 
and radiating like a black body one finds a luminosity of 
\begin{equation}
L_{\nu} = 2.4 \times 10^{-18} R_S^2 cos(i) T^{8/3} \nu^{1/3} (erg/s)
\end{equation}
(Peterson 1997, see also Ishibashi \&  Courvoisier 2009, Kishimoto et al. 2008).
At infrared wavelengths and shorter the spectrum is dominated by the emission 
of the inner disk radii. 
In the NIR Ks-band 1~mJy corresponds to 3 L$_{\odot}$.
For $cos(i)$$\sim$1, a Schwarzschild radius of $R_S$=10$^{12}$cm,
and an observing frequency of 136.364~THz for a wavelength of 2.2$\mu$m we obtain
$L_{\nu} = 1.235 \times 10^{11} T^{8/3}(erg/s)$.
So a minimum disk temperature of $T=2.6 \times 10^8K$ is required
to obtain $L_{\nu} = 1 L_{\odot}$.
For higher temperatures one needs to introduce a surface filling
factor of less than unity to explain the Ks-band emission in this model.
However, one requires a high number density to have an optically thick disk.
In addition the $\nu^{1/3}$ dependency is in contradiction with the
steep infrared spectra that have also been observed during 
flares
(Ghez et al. 2005ab, Eisenhauer et al. 2005, Gillessen et al. 2006, 
Hornstein et al. 2006, Krabbe et al. 2006).
These facts make a geometrically thin and optically thick disk 
a very unlikely dominant contributer to explain a possible quiescent Ks-band 
luminosity of SgrA*.

\section{Summary and conclusion}
\label{section:Summary}

We have discussed the variable infrared and X-ray emission of SgrA*
in its extreme luminosity states.
We reported on Ks-band imaging of the 
SMBH counter part of SgrA* in its low state
and investigated the structure and brightness of the central 
S-star cluster that surrounds the SMBH at the position of SgrA*.
We have used three independent methods to remove or strongly suppress the
flux density contributions of stars in the central 2'' diameter region 
around SgrA*. 
All methods revealed faint extended emission around the SgrA* position.
De-reddened with $A_K$=2.8 the peak emission of the extended S-star cluster is 
about 1.3~mJy within one resolution element, 
corresponding to a surface flux density of about 0.5~Jy (dereddened) per square arcsecond. 
For the dataset taken
on 23 September 2004 no NIR counterpart of SgrA* was detected with a
flux density limit of about 2~mJy (dereddened).
For 30 August 2004 we find a NIR counterpart of SgrA* with a
dereddened flux density of about 4~mJy.
We find that the luminosity during the low state can most likely be accounted 
for by synchrotron emission from a continuous or spotted accretion disk.
In this case we expect the possible quiescent source associated with SgrA* to be
significantly polarized.

Steep spectral indices observed in the NIR wavelength domain may 
reflect the presence of typical cutoff frequencies.
In addition to the cooling break $\nu_{break}$ 
as discussed in Sect.~\ref{subsubsection:Essentials}
there may be a cutoff in the relativistic electron spectrum at work
(e.g. Eckart et al. 2006a, Liu et al. 2006).
This will result in a modulation of the intrinsically flat spectra
with an exponential cutoff proportional to $exp[-(\lambda_0/\lambda)^{0.5}]$
(see e.g. Bregman 1985, and Bogdan \& Schlickeiser 1985)
and a cutoff wavelength $\lambda_0$ in the infrared.
If $\lambda_0$ lies in the 4-8$\mu$m wavelength range (Eckart et al. 2006a),
the variation in the spectral
index is on the order of $\Delta \alpha$$\sim$1.0.
In a number of extragalactic jets 
these cutoffs have been observed to be relevant in the NIR wavelength domain
(3C~293: Floyd et al. 2006,
M87: Perlman et al. 2001; Perlman \& Wilson 2005;
3C~273: Jester et al. 2001, 2005). 
In these sources the cutoff frequency at which synchrotron losses become dominant
is $\nu_{jet,break}$=4$\times$10$^{13}$Hz (or 7.5$\mu$m wavelength),
which is remarkably similar to what may be required in the case of 
the Galactic center (Eckart et al. 2006a).
For SgrA* such a cutoff can quite naturally explain the steep NIR spectral indices reported by 
(Eisenhauer et al. 2005, Gillessen et al. 2006, Krabbe et al. 2006) despite the
low upper flux density limits in the MIR.

For the three brightest X-ray flares the SSC emission from THz peaked source components
can fully account for the observed flux density variations observed in the
NIR and X-ray domain.
Here models in which the MIR/NIR flux density contributions are due
to the high frequency tail of the associated synchrotron spectrum are favored 
(models A in Tables~\ref{modeldata} and~\ref{modeldensity}).
Combined with the source component size of 0.2-0.3~$R_0$ and relativistic expansion 
within or towards the end of the flare, the spectral energy densities of
relativistic electrons are compatible with the electron number density derived for the 
accretion flow at 10~R$_s$ distance 
and the estimated upper limits of the mass accretion rate onto SgrA*.
For the weak X-ray flare first discussed by Eckart et al. (2004) these boundary conditions 
are fulfilled for both model approaches (A and B).
For model A$_{low}$ we expect significant polarization of the NIR flux density, 
whereas for model B$_{low}$ the scattering process should lead to a significantly 
lower degree of polarization.

For the 4 April 2007 flare our favored model A$\gamma$ of the peak fluxes 
meets the 11.8$\mu$m 3$\sigma$ limit and is only by 1$\sigma$ off the expected 
flux density at 3.8$\mu$m wavelength. In the X-ray domain it matches the mean flux exactly and the 
spectral slope by 0.8$\sigma$. Giving these four quantities the same weight we obtain
a reduced $\chi^2_{red}$ value of $\chi^2_{red}$=10.7/d.o.f=3.7.
The X-ray data cover a comparatively large and rather densely sampled spectral range.
We get $\chi^2_{red}$$\le$2
if the X-ray information is weighted higher by a factor of two or more 
(see Porquet et al. 2008, Dodds-Eden et al. 2009).
A~~$\chi^2$-based comparison of different single component emission models 
over a large wavelength range (radio or infrared till X-ray) is, however, 
only of limited value.
More adequate physical models of the compact emission from SgrA*
contain on some level contributions from several source components, like
the hot-spot/disk (Eckart et al. 2006b, Broderick \& Loeb 2006)
or multi-spot models (Eckart et al. 2008a , see also Pech\'a\v{c}ek et al., 2008),
the spiral arm models (Karas et al. 2007, Falanga et al. 2007),
the more elaborate evolving hot-spot model by (Yuan et al. 2008),
or the jet/jet-base models
(Falcke \& Markoff 2000, 2001, Markoff 2005, Markoff et al. 2007).
This implies that in addition to a possible dominant emission component, 
secondary source or spectral components will contribute to the flux 
and spectral index information in 
the different wavelength bands across the electromagnetic spectrum. 
This complicates the comparison of different models.
Distinguishing between emission mechanisms 
requires more work on detailed flare modeling, using time dependent
dense flux density sampling across a broad frequency band, preferentially including
polarization and spatial information.

Here we presented a new SSC model for the flare event on April 4, 2007
that accounts for the available simultaneous peak flux density and spectral 
index information. 
Variations in the SSC scattering explain 
that the flare profile in the X-ray domain is narrower than that 
at NIR wavelengths. 
This suggests at least for this flare event that models in which the 
NIR flares are due to optically thin synchrotron radiation rather 
than inverse Compton scattered light are preferred.
In general models in which the MIR/NIR flux densities are explained through the low frequency 
tail of the inverse Compton scattered spectrum are disfavored.
They result in higher masses for the emitting sources.
 
We conclude that both weak and bright flares from SgrA* can be explained by 
the synchrotron self-Compton mechanism. 
This emission mechanism may explain
the flare activity over up to two orders in magnitude in the NIR or
X-ray flux density and over six orders of magnitude in frequency.
No other dominant emission mechanisms are required.
Densities of relativistic electrons of a few times 10$^7$ cm$^-3$
(comparable to what is derived from X-ray data on a 1000 Schwarzschild
radii scale) are also required close to the SgrA* black hole to
explain the flare emission.

The variable synchrotron emission 
may be limited at short wavelengths by a NIR cutoff (cooling break) due to 
synchrotron losses.
The self-Compton scattered spectrum may be limited at the longest wavelengths by the 
low energy cutoff in the relativistic electron distribution.
If future sub-mm and X-ray obervations allow for a simultaneous detection of
the THz turnover frequency ($\nu_m$) and a low energy cutoff of the SSC spectrum
($\nu_{SSC,upper}$),
information on the low energy
cutoff in the relativistic electron distribution can be obtained via
$\gamma$$_1$$^2$$\sim$$\nu_{SSC,upper}$$\nu_m^{-1}$.
In a few years further polarimetric NIR observations
at high angular resolution and sensitivity
will help to distinguish between the importance of the different
emission mechanisms in the low state of SgrA*.
Interferometric observations in the near-infrared using GRAVITY
at the VLTI (Eisenhauer et al. 2008, Straubmeier et al. 2008, Zamaninasab et al. 2009)
and in the mm-wavelength domain
(e.g. Fish et al. 2009, Doeleman et al. 2009)
will be important to understand the emission from SgrA*.

\begin{acknowledgements}
N. Sabha is member of the Bonn Cologne Graduate School for
Physics and Astronomy.
This work was supported in part by the Deutsche Forschungsgemeinschaft
(DFG) via grant SFB 494, the Max Planck Society through 
the International Max Planck Research School, as well as 
special funds through the University of Cologne.
We are grateful to all members
of the NAOS/CONICA and the ESO PARANAL team.  
Macarena Garc\'{\i}a-Mar\'{\i}n is supported by the German 
federal department for education and research (BMBF) under 
the project numbers 50OS0502 \& 50OS0801.
M. Zamaninasab and D. Kunneriath
are members of the International Max Planck Research School (IMPRS) for
Astronomy and Astrophysics at the MPIfR and the Universities of
Bonn and Cologne. 
R. Sch\"odel acknowledges support by the Ram\'on y Cajal
programme by the Ministerio de Ciencia e Innovaci\'on of the
government of Spain.

\end{acknowledgements}

\vspace*{0.5cm}

\rf{Alexander, T.; Hopman, C., 2009, ApJ 697, 1861}

\rf{Bondi, H., 1952, MNRAS 112, 195}

\rf{Baganoff, F.K., 
   Bautz, M.W., Brandt, W.N., et al. 2001, Nature, 413, 45}

\rf{Baganoff, F. K., et al., 2002, 201st AAS
  Meeting, \#31.08; Bulletin of the American Astronomical Society,
  Vol. 34, 1153}

\rf{Baganoff, F. K., Maeda,
  Y., Morris, M., et al. 2003, ApJ 591, 891}

\rf{Bahcall, J. N.; Wolf, R. A., 1976, ApJ 209, 214}

\rf{B\'elanger, G.; Terrier, R.; de Jager, O. C.; Goldwurm, A.; Melia, F., 2006, JPhCS 54, 420}

\rf{Blum, R. D.; Sellgren, K.; Depoy, D. L., 1996, ApJ 470, 864}

\rf{Bogdan, T.J., \& Schlickeiser, R., 1985, \aa, 143, 23}

\rf{{Bower}, G.~C., {Falcke}, H., {Herrnstein}, R.~M., {Zhao}, J., {Goss}, W.~M.,
  \& {Backer}, D.~C. 2004, Science, 304, 704}

\rf{Bregman, J.N., 1985, \aj, 288, 32}

\rf{Broderick, A.E., Loeb, A., 2006, MNRAS 367, 905}

\rf{Buchholz, R. M.; Sch\"odel, R.; Eckart, A, 2009, A\&A 499, 483}

\rf{Cuadra, J.; Nayakshin, S., 2006, JPhCS 54, 436}

\rf{Cuadra, J.; Nayakshin, S.; Martins, F., 2008, MNRAS 383, 458}

\rf{Diolaiti, E. Bendinelli, O., Bonaccini, D., Close, L., Currie, D., Parmeggiani, G., A\&AS 147, 335-346, 2000}

\rf{Do, T.; Ghez, A.M.; Morris, M.R.; Yelda, S.; Meyer, L.; 
     Lu, J.R.; Hornstein, S.D.; Matthews, K., 2009a, ApJ 691, 1021.}

\rf{Do, T.; Ghez, A.M.; Morris, M.R.; Lu, J.R.; Matthews, K.; 
        Yelda, S.; Larkin, J., 2009b, 2009arXiv0908.0311D	}

\rf{Doeleman, S.S.; Fish, V.L.; Broderick, A.E.; Loeb, A.; Rogers, A.E.E., 2009, ApJ 695, 59D}

\rf{Dodds-Eden, K.; Porquet, D.; Trap, G.; Quataert, E.; Haubois, X.; 
        Gillessen, S.; and 19 coauthors, 2009, ApJ 698, 676	}

\rf{Eckart, A.; Duhoux, P. R. M. 1990, ASPC, 14, 336}

\rf{Eckart, A. \& Genzel, R. 1996, Nature 383, 415}

\rf{Eckart, A., Genzel, R., Ott, T. and Sch\"odel, R. 2002, MNRAS, 331, 917}

\rf{Eckart, A.; Moultaka, J.; Viehmann, T.; Straubmeier, C.; Mouawad, N.; 
        Genzel, R.; Ott, T.; Sch\"odel, R.; Baganoff, F. K.; Morris, M. R, 2003, ANS 324, 557}

\rf{Eckart, A.; Baganoff, F. K.; Morris, M.; and 11 coauthors, 2004, A\&A 427, 1}

\rf{Eckart, A.; Baganoff, F. K.; Sch\"odel, R.; Morris, M.; Genzel, R.; Bower, G. C.; Marrone, D.; et al.  2006a, A\&A 450, 535}

\rf{Eckart, A.; Sch\"odel, R.; Meyer, L.; Trippe, S.; Ott, T.; Genzel, R.,  2006b, A\&A 455, 1}

\rf{Eckart, A.; Baganoff, F. K.; Zamaninasab, M.; Morris, M. R.; 
  Sch\"odel, R.; and 16 coauthors, 2008a, A\&A 479, 625}

\rf{Eckart, A.; Sch\"odel, R.; 
    Garcia-Marin, M.; Witzel, G.; Weiss, A.; Baganoff, F. K.; Morris, M. R.; Bertram, T.; 
    Dovciak, M.; Duschl, W. J.; and 21 coauthors, 2008b, A\&A 492, 337}

\rf{Eckart, A.; Baganoff, F. K.; Morris, M. R.; Kunneriath, D.; 
    Zamaninasab, M.; Witzel, G.; Sch\"odel, R.; Garcia-Marin, M.; Meyer, L.; Bower, G. C.; 
    and 10 coauthors, 2009, A\&A 500, 935.}

\rf{Eisenhauer, F.; Sch\"odel, R.; Genzel, R.; Ott, T.;
  Tecza, M.; Abuter, R.; Eckart, A.; Alexander, T., 2003, ApJ 597, L121}

\rf{Eisenhauer, F.; Genzel, R.; Alexander, T.; 
   Abuter, R.; Paumard, T.; Ott, T.; ; and 15 coauthors, 2005, ApJ 628, 246}

\rf{Eisenhauer, F.; Perrin, G.; Brandner, W.; Straubmeier, C.; 
   Richichi, A.; Gillessen, S.; Berger, J. P.; Hippler, S.; Eckart, A.; 
   Sch\"oller, M.; and 38 coauthors, 2008, SPIE 7013, 69}

\rf{Fish, V.L.; Doeleman, S.S.; Broderick, A.E.; Loeb, A.; Rogers, A.E.E., 2009, ApJ 706, 1353}

\rf{Falanga, M., Melia, F., Tagger, M., Goldwurm, A., Belanger, G., 2007, ApJ 662, L15}

\rf{Falcke, H.; Markoff, S.; Fender, R.,       2001, AGM 18, 107}

\rf{Falcke, H.; Markoff, S., 2000, A\&A 362, 113}

\rf{Floyd D.J.E., Eric Perlman, E., Leahy, J.P., and 5 coauthors, 2006, ApJ 639, 23}

\rf{Genzel, T., Eckart, A., Ott, T. \& Eisenhauer,
   MNRAS 1997, 291, 219}

\rf{Genzel, R., Pichon, C., Eckart, A., Gerhard,
   O.E., Ott, T. 2000, MNRAS 317, 348}

\rf{Genzel, R., Sch\"odel, R., Ott, T., et al. 2003, Nature, 425, 934}

\rf{Gezari, S., Ghez, A.~M., Becklin, E.~E., Larkin, J., 
        McLean, I.~S., Morris, M., ApJ 576, 790, 2002}

\rf{Ghez, A., Klein, B.L., Morris, M. \& Becklin,
   E.E. 1998, ApJ, 509, 678}

\rf{Ghez, A., Morris, M., Becklin, E.E., Tanner,
   A. \& Kremenek, T.  2000, Nature 407, 349}

\rf{Ghez, A. M., Duch\'ene, G., Matthews, K., et
  al. 2003, ApJ, 586, L127}

\rf{Ghez, A.M., Wright, S.A., Matthews, K., et
    al. 2004a, ApJ 601, 159}

\rf{Ghez, A.M., Hornstein, S.D., Bouchez, A., Le Mignant, D., Lu, J., 
   Matthews, K., Morris, M., Wizinowich, P., Becklin, E.E., 2004b, AAS 205, 2406}

\rf{Ghez, A.M., Salim, S., Hornstein, S. D., 
   Tanner, A., Lu, J. R., Morris, M., Becklin, E. E., Duch\^ene, G., 2005a, ApJ 620, 744	}

\rf{Ghez, A. M.; Hornstein, S. D.; Lu, J. R.; Bouchez, A.; Le Mignant, D.; van Dam, M. A.; 
        Wizinowich, P.; Matthews, K.; Morris, M.; Becklin, E. E.; and 7 coauthors, 2005b, ApJ 635, 1087}

\rf{Gillessen, S.; Eisenhauer, F.; Trippe, S.; Alexander, T.; Genzel, R.; Martins, F.; Ott, T.,
         2009, ApJ 692, 1075}

\rf{Gillessen, S.; Eisenhauer, F.; Quataert, E.; Genzel, R.; 
    Paumard, T.; Trippe, S.; Ott, T.; Abuter, R.; Eckart, A.; 
    and 4 coauthors, 2006, ApJ 640, L163	}

\rf{Goldwurm, A., Brion, E.,
  Goldoni, P. et al. 2003, ApJ, 584, 751}

\rf{Gould, R.J., 1979, A\&A 76, 306}

\rf{Hawley, J.F.; Balbus, S.A., 
   1991, ApJ 376, 223}

\rf{Hawley, J.F.; Stone, J.M., 1998, ApJ 501, 758, }

\rf{Hornstein, S.D.; Ghez, A.M.; Tanner, A.; Morris, M.; Becklin, E. E.; Wizinowich, P., 2002, ApJ 577, L9}

\rf{Hornstein, S.D., et al., 2006, JPhCS 54, 399}

\rf{Hornstein, S. D., Matthews, K., Ghez, A. M., Lu, J. R., Morris, M., Becklin, E. E., Rafelski, M., \& Baganoff, F. K. 2007, ApJ, 667, 900}

\rf{Igumenshchev, I.V., 2002, ApJ 577, 31}

\rf{Ishibashi, W.; Courvoisier, T. J.-L., 2009, A\&A 495, 113}

\rf{Jester, S., R\"oser, H.-J., Meisenheimer, K., \& Perley, R. 2005, A\&A, 431, 477}

\rf{Jester, S., R\"oser, H.-J., Meisenheimer, K., Perley, R., \& Conway, R. 2001, A\&A, 373, 447}

\rf{Karas, V., Dovciak, M., Eckart, A., Meyer, L., 2007, Proceedings of 
   the Workshop on the Black Holes and Neutron Stars, eds. S. Hledik and Z. Stuchlik, 
   19-21 September 2007 (Silesian University, Opava), pp. 99-108, arXiv:0709.3836}

\rf{Kardashev, N.S., 1962, SvA 6, 317}

\rf{Kishimoto, M.; Antonucci, R.; Blaes, O.; Lawrence, A.; Boisson, C.; Albrecht, M.; Leipski, C., 2008, Nature 454, 492}

\rf{Krabbe, A.; Iserlohe, C.; Larkin, J. E.; Barczys, M.; McElwain, M.; 
           Weiss, J.; Wright, S. A.; Quirrenbach, A.  2006, ApJ 642, L145}

\rf{Lightman, A. P., \& Shapiro, S. L. 1977, ApJ, 211, 244 }

\rf{Liu, S.; Petrosian, V.; Melia, F; Fryer, C, 2006, ApJ 648, 1020}

\rf{Lucy, L. B. 1974, AJ, 79, L745}

\rf{Mahadevan, R. \& Quataert, E. 1997, ApJ, 490, 605}

\rf{Markoff, S., Falcke, H., Yuan, F. \& Biermann, P.L.
   2001, A\&A, 379, L13}

\rf{Markoff, S., 2005, ApJ 618, L103 }

\rf{Markoff, S.; Bower, G.C.; Falcke, H., 2007, MNRAS 379, 1519}

\rf{Marrone, D.P.; Moran, J.M.; Zhao, J.-H.; Rao, R. 2007, ApJ 654, L57}

\rf{Marrone, D. P.; Baganoff, F. K.; 
    Morris, M.; and 14 coauthors, 2008, ApJ 682, 373}

\rf{Marscher, A.P. 1983, ApJ, 264, 296}

\rf{Melia, F. \& Falcke, H. 2001, ARA\&A 39, 309}

\rf{Meyer, L., Eckart, A., Sch\"odel, R., Dovciak, M., Karas, V., Duschl, W.J., 2007, A\&A 473, 707 }

\rf{Meyer, L., Eckart, A., Sch\"odel, R., Duschl, W. J., Muciz, K., Dovciak, M., Karas, V., 2006a, A\&A 460, 15}

\rf{Meyer, L., Sch\"odel, R., Eckart, A., Karas, V., Dovciak, M., Duschl, W. J., 2006b, A\&A 458, L25}

\rf{Murphy, B. W., Cohn, H. N., \& Durisen, R. H. 1991, ApJ, 370, 60}

\rf{Nayakshin, S.; Sunyaev, R., 2003, MNRAS 343, L15}

\rf{\"Ozel, F., Psaltis, D., \& Narayan, R. 2000, ApJ, 541, 234}

\rf{Ott, T.; Eckart, A.; Genzel, R., 1999, ApJ 523, 248	}

\rf{Page, K.L.; Reeves, J.N.; O'Brien, P.T.; Turner, M.J. L.; Worrall, D.M., 2004, MNRAS 353, 133 }

\rf{Pech\'a\v{c}ek, T., Karas, V., Czerny, B., 2008, A\&A 487, 815}

\rf{Perlman, E. S., Biretta, J. A., Sparks, W. B., Macchetto, F. D., \& Leahy, J. P.  2001, ApJ, 551, 206}

\rf{Perlman, E. S., \& Wilson, A. S. 2005, ApJ, 627, 140}

\rf{Peterson, B.M., 1997, 'An introduction to active galactic nuclei', p. 44, Cambridge University Press}

\rf{Porquet, D., Predehl, P., Aschenbach, et al.
    2003, A\&A 407, L17}

\rf{Porquet, D.; Grosso, N.; Predehl, P.; Hasinger, G.; Yusef-Zadeh, F.; 
    and 11 coauthors; 2008, A\&A 488, 549}

\rf{Quataert, E. 2002, ApJ, 575, 855}

\rf{Quataert, E., Astron. Nachr., Vol. 324,
  No. S1 (2003), Special Supplement "The central 300 parsecs of the
  Milky Way", Eds. A.  Cotera, H. Falcke, T. R. Geballe, S. Markoff,
  p. 435 (astro-ph/0304099)}

\rf{Reid, Mark J., 1993, ARA\&A 31, 345}

\rf{Straubmeier, Christian; Eisenhauer, Frank; Perrin, Guy; 
    Brandner, Wolfgang; Eckart, Andreas, 2008, SPIE 7013, 93}

\rf{Sch\"odel, R.; Merritt, D.; Eckart, A., 2009, A\&A 502 91}

\rf{Sch\"odel, R.; Eckart, A.; Alexander, T.; Merritt, D.; Genzel, R.; 
         Sternberg, A.; Meyer, L.; Kul, F.; Moultaka, J.; Ott, T.; Straubmeier, C.  2007, A\&A 469, 125}

\rf{Sch\"odel, R., Genzel, R., Ott, et al. 2003,
  ApJ, 596, 1015}

\rf{Sch\"odel, R., Ott, T., Genzel, R. et
  al. 2002, Nature, 419, 694}

\rf{Scoville, N. Z.; Stolovy, S. R.; Rieke, M.; Christopher, M.; Yusef-Zadeh, F., 2003, ApJ 594, 294}

\rf{Tanner, A.; Ghez, A. M.; Morris, M.; Becklin, E. E.; Cotera, A.; and 3 coauthors, 2002, ApJ 575, 860	}

\rf{Tanner, A.; Ghez, A. M.; Morris, M. R.; Christou, J. C., 2005, ApJ 624, 742}

\rf{van der Laan, H., 1966, Nature 211, 1131}

\rf{Yuan, F., Quataert, E., \& Narayan, R. 2003, ApJ, 598, 301}

\rf{Yuan, F., 2006, JPhCS 54, 427}

\rf{Yuan, F., Lin, J.; Wu, K.; Ho, Luis C., 2008, MNRAS 395, 2183}

\rf{Yusef-Zadeh, F., Roberts, D., Wardle, M., Heinke, C. O., Bower, G. C., 2006a,  ApJ 650, 189}

\rf{Yusef-Zadeh, F., et al., 2006b, ApJ 644, 198}

\rf{Yusef-Zadeh, F.; Wardle, M.; Cotton, W. D.; Heinke, C. O.; Roberts, D. A., 2007, ApJ 668, L47}

\rf{Yusef-Zadeh, F.; Wardle, M.; Heinke, C.; Dowell, C. D.; 
     Roberts, D.; Baganoff, F. K.; Bower, G. C., 2008, ApJ 682, 361}

\rf{Yusef-Zadeh, F.; Bushouse, H.; Wardle, M.; Heinke, C.; Roberts, D. A.; 
    and 16 coauthors, 2009, 2009arXiv0907.3786Y	}

\rf{Zamaninasab, M., Eckart, A., Meyer, L., Sch\"odel, R., Dovciak, M., Karas, V.,
    Kunneriath, D., Witzel, G., Giessuebel, R., K\"onig, S., Straubmeier, C., Zensus, A.,
    2008,
    Proc. of a conference on 'Astrophysics at High Angular Resolution (AHAR 08)'
    held 21-25 April 2008 in Bad Honnef, Germany, 2008, JPhCS 131, 12008}

\rf{Zamaninasab, M.; Eckart, A.; Witzel, G.; Dovciak, M.; Karas, V.; Giessuebel, R. Sch\"odel R.;
Bremer, M.; Garcia-Marin, M.; Kunneriath, D.; Muzic, K.; and 4 coauthors
2009, arXiv0911.4659, A\&A in press}
                                          
\rf{Zhao, J.-H., Young, K.H.,  Herrnstein, R.M., Ho, P.T.P., Tsutsumi, T., Lo,
  K.Y., Goss, W.M. \& Bower, G.C., 2003, ApJL, 586, L29.}

\end{document}